\documentclass[english,aps,pre,superscriptaddress,twocolumn,showpacs]{revtex4-1}
\usepackage{graphicx,amsfonts,amsmath,amssymb,color}
\usepackage[english]{babel}

\begin{document}

\title{Event-triggered feedback in noise-driven phase oscillators}

\author{Justus A. Kromer}
\email{justuskr@physik.hu-berlin.de}
\affiliation{Department of Physics, Humboldt-Universit\"{a}t zu Berlin, 
Newtonstr. 15, 12489 Berlin, Germany}
\author{Benjamin Lindner}
\affiliation{Department of Physics, Humboldt-Universit\"{a}t zu Berlin, 
Newtonstr. 15, 12489 Berlin, Germany}
\affiliation{Bernstein Center for Computational Neuroscience Berlin, Germany}
\author{Lutz Schimansky-Geier}
\affiliation{Department of Physics, Humboldt-Universit\"{a}t zu Berlin, 
Newtonstr. 15, 12489 Berlin, Germany}
\affiliation{Bernstein Center for Computational Neuroscience Berlin, Germany}

\begin{abstract}
Using a stochastic nonlinear phase oscillator model, we study the effect of event-triggered feedback on the statistics of interevent intervals. Events are associated with the entering of a new cycle. The feedback is modeled
by an instantaneous increase (positive feedback) or decrease (negative feedback) of the oscillators frequency, whenever an event occurs followed by an exponential decay on a slow timescale. 
In contrast to previous works, we also consider positive feedback that leads to various novel effects. For instance, besides the known excitable and oscillatory regime, that are separated by a saddle-node on invariant circle bifurcation, positive feedback can lead to bistable dynamics and a change of the system's excitability.
The feedback has also a strong effect on noise-induced phenomena like coherence resonance or anti-coherence resonance.
Both positive and negative feedback can lead to more regular 
output for particular noise strengths. Finally, we investigate serial correlation in the sequence of interevent intervals that occur due to the additional slow dynamics. 
We derive approximations for the serial correlation coefficient and show that positive feedback results in extended positive interval correlations whereas negative feedback yields short-ranging negative correlations. Investigating the interplay of 
feedback and the nonlinear phase dynamics close to the bifurcation, we find that correlations are most pronounced for an optimal feedback strengths.
\end{abstract}

\maketitle

\section{Introduction}
Self-sustained oscillations occur in many physical, chemical or biological systems \cite{ermentrout1984beyond}. 
If variations of the amplitude are negligible, a widely-used model in this context is the well-known dynamics for the phase $\phi$ \cite{kuramoto1984chemical}:
\begin{align}
\label{equ.noFeedbackCompleteSystem}
 \dot{\phi}(t) =& \omega_0 - \epsilon \sin[\phi (t)].
\end{align}
Here $\omega_0$ represents the oscillators frequency in the case $\epsilon \rightarrow 0$. Without loss of generality, we restrict our investigations on $\omega_0>0$. By rescaling $\omega_0$ and the timescale, $\epsilon$ can be set to one (dimensionless units). 
The system can show both excitable ($0 < \omega_0 < 1$) or oscillatory ($\omega_0 > 1$) dynamics. Both regimes are separated by a saddle-node on invariant circle (SNIC) bifurcation at $\omega_0=1$, which makes
the system to a good model for class I excitability \cite{lindner2004effects}. 
Eq. (\ref{equ.noFeedbackCompleteSystem}) is known as the Adler's equation \cite{adler1946study} and is often used to describe excitability in 
optical system \cite{giudici1997andronov, goulding2007excitability} or in neuroscience \cite{izhikevich2007dynamical}, 
particle motion in a tilted periodic potential, or 
to study the onset of resistance in superconducting Josephson junctions \cite{stewart1968current,strogatz1994nonlinear}.
Generally, such oscillators are studied
when driven by time-dependent forces, such as noise, when subjected to time delayed feedback \cite{aust2010delay}, or when they are coupled in networks.

For many applications, particular events in the phase dynamics are of foremost interest, e.g. the crossings of a threshold value
$\phi=2 \pi$ as, for instance, associated with the generation of an action potential in a nerve cell, the dropout of 
light intensity in an excitable laser, the release of a messenger by a cell, or the division of a cell. The statistics of the intervals between these events $\Delta t$ (interevent intervals or in the following IEI) in the presence of noise have been studied intensely in the neurobiological context for the related class
of integrate-and-fire models \cite{FouBru02, burkitt2006} (here IEIs are referred to as interspike intervals). 

In some systems, the events directly influence the dynamics of the oscillator. Put differently, in these systems we find
event-triggered feedback mechanisms.
Generally, the oscillator's dynamics becomes more interesting if such feedback mechanisms are taken into account. For neurons negative feedback can arise from slow inhibitory ionic currents that change over several IEIs. 
This can lead to spike-frequency adaptation \cite{LiuWan01, benda2003universal}, noise shaping \cite{ChaLin05}, and interval correlations \cite{SchFis10, AviCha11}. 
Feedback, however, can be also positive, for instance, due to variations in the external potassium concentration,
which are triggered by neural spiking \cite{strauss2008increasing, postnov2009dynamical} and act on a timescale which is large compared to the individual IEIs \cite{frohlich2008potassium}. In some systems strong positive feedback can change the dynamics fundamentally, leading, for instance, to bursting behavior \cite{frohlich2006slow}. 
In laser physics 
positive feedback for particular modes can be used to self-mode-lock lasers \cite{derickson1990self} and it seems to be a plausible explanation of positive IEI correlations, reported in Ref. \cite{schwalger2012interspike}. 
In cell biology, positive feedback loops occur, for instance, in the lactose utilization network of the Escherichia coli, where the production of lactose permease increases its expression level and is assumed to be a reason for 
bistability in the lactose utilization \cite{ozbudak2004multistability, choi2008stochastic}. 
However, the effect of positive feedback, especially in the presence of noise, is so far only poorly understood.

     Analytical attempts to deal with an 
additional feedback dynamics in a pulse     
     generator were mainly limited to 
approximations of the firing rate 
\cite{ermentrout1998linearization, LiuWan01}
and weak feedback approximations
for the IEI statistics of a very simple integrate-and-fire
model, the so-called perfect IF model
\cite{urdapilleta2011survival}
.
Regarding the more striking 
feature of the feedback-induced 
interspike interval correlations, 
approximations until recently were carried 
 out for the perfect IF model
\cite{SchFis10}, variants that 
deviate only by a weak nonlinearity from 
it \cite{schwalger2013leak}, or integrate-and-fire models subjected to a weak feedback \cite{Urd11}. In \cite{schwalger2013patterns}, a 
general theory has been worked out to 
calculate patterns of interval 
correlations in multidimensional IF 
models. All these studies focused on a 
negative feedback, however, and did not address the generic phase oscillator dynamics eq. (\ref{equ.noFeedbackCompleteSystem}).

Here we study the dynamics of a phase oscillator in the vicinity of a saddle-node on invariant circle bifurcation from the excitable to the oscillatory regime, which is subject to noise and an event-triggered feedback. 
We consider feedback strengths that can attain both positive or negative values and derive analytic approximation for several statistical measures by considering a large timescale separation between the
phase and the feedback dynamics. 

Our results for negative feedback are in line
with previous studies: we find suppression of low-frequency power in the power spectrum of the spike train \cite{chacron2004noise,ChaLin07} and negative serial correlations 
in the series of $N$ subsequent
IEIs $\Delta t_1, ... , \Delta t_N$, \cite{LiuWan01,ChaLon00,PreSej08, farkhooi2009serial}.
More remarkably, we find that positive feedback causes a number of novel effects. In the deterministic system, bistability emerges in the form of the coexistence of a stable node (SN) and a limit cycle (LC) attractor. 
Secondly, we study the effect of noise and feedback on the system. Here we focus on the excitable and the oscillatory regime.
We find anti-coherence resonance in the excitable regime - IEI variability is maximized at a finite noise intensity - and observe positive IEI correlations in both, the excitable and the oscillatory regime.
Interestingly, IEI correlations for both positive and negative feedback behave non-monotonically with the feedback strength, if the system is close to the bifurcation.

Our paper is organized as follows. In section \ref{sec:Model} we introduce the model and the statistics of interest. We study first, in section \ref{sec:cyclePeriod}, the non-linear dynamics of the system 
without noise (including a bifurcation analysis) and explore the effects of noise and feedback on the mean frequency of the oscillator. In section \ref{sec:IEIVariability}, we investigate 
the IEI variability and the power spectrum of the phase oscillator with feedback. Section \ref{sec:SerialCorrelations} is devoted to IEI correlations. 
Finally, we conclude by summarizing our results and discussing their broader implications. All details concerning simulation techniques and analytical calculations of the serial correlation coefficient are
given in Appendix \ref{sec:simulationtechniques} and \ref{sec:SerialCorrelationsPRC}, respectively.


\section{The Model}
\label{sec:Model}
\begin{figure}[t]
\begin{minipage}[t]{0.7\linewidth}
\centering 
   \includegraphics[width=1\linewidth]{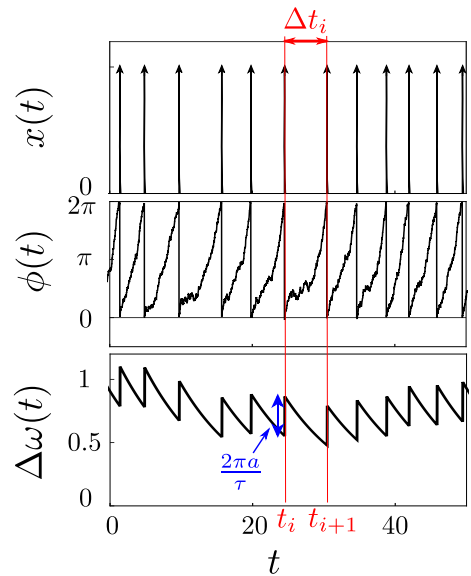}
\end{minipage}
  \caption{(Color online) Time evolution of $x(t)$, $\phi(t)$, and $\Delta \omega(t)$ (from top to bottom) for $\omega_0= 1.1$,
$a=0.5$, $\tau=10$, and $D=0.1$.}
 \label{fig:Trajektories1}
 \end{figure} 


In order to implement the feedback we define an event to occur whenever the phase reaches the threshold $2 \pi$, i.e., $\phi(t_i)=2 \pi$, where $t_i$ denotes the time of the $i$th event. Afterwards, the phase is reset $(\phi \rightarrow 0)$.
The feedback acts on the phase oscillator by increasing (positive feedback) or reducing (negative feedback) its frequency.
Thus, we add a time-dependent part $\Delta \omega(t)$ to the frequency $\omega_0$, which accounts for the frequency adaptation due to the feedback. Consequently, \mbox{eq. (\ref{equ.noFeedbackCompleteSystem})} becomes
\begin{align}
\label{equ.CompleteSystem}
\begin{array}{rl}
 \dot{\phi}(t) =& \Delta \omega(t) + \omega_{0} - \sin[\phi (t)] + \sqrt{2 D} \xi(t)\\
\end{array}.
\end{align}
Combined with the reset condition
\begin{align}
\label{equ.reset}
\begin{array}{rl}
 \text{if} \ \  \phi = 2 \pi & \text{\ , \ then} \ \ \phi \rightarrow 0 \\
\end{array}.
\end{align}
Here we also added white Gaussian noise [$\langle \xi(t) \rangle=0$ and $\langle \xi(t)\xi(t') \rangle=\delta(t-t')$] with a noise strength $D$.
Where $\langle . \rangle$ denotes averaging.

When an event occurs, the system perceives a kick which changes $\Delta \omega$. This is modeled by the additional dynamics
\begin{align}
\label{equ.TriggeredFeedback}
\tau \frac{d}{dt}\Delta \omega(t) = - \Delta \omega(t)+ 2 \pi a \ x(t),
\end{align}
where
\begin{align}
\label{equ:outputFunction}
 x(t)=\sum \limits_{i} \delta(t-t_{i})
\end{align}
is the sequence of kicks at the event times $t_i$.

 \begin{figure}[t]
\begin{minipage}[t]{1\linewidth}
\centering 
   \includegraphics[width=1\linewidth]{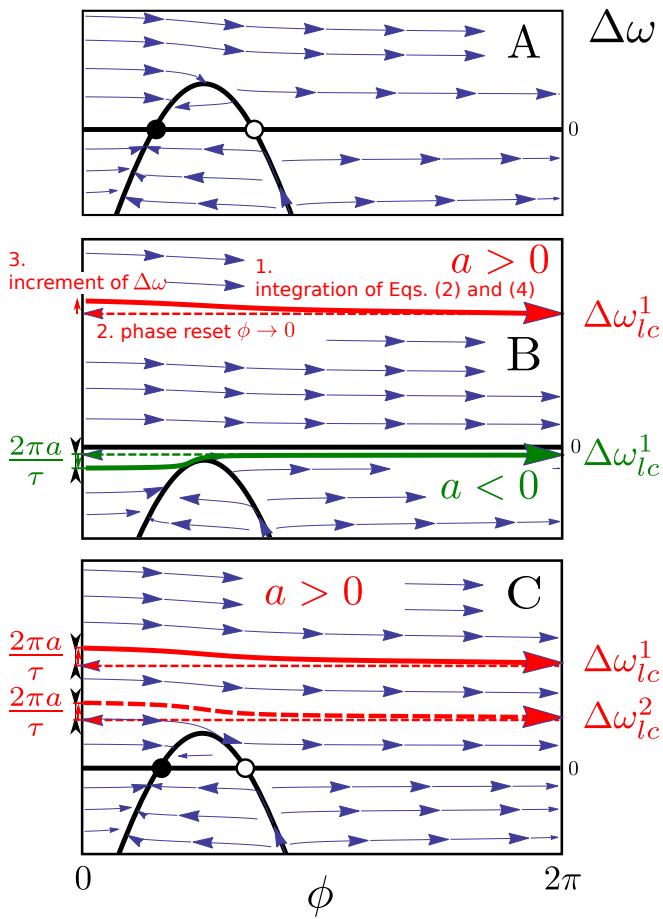}
\end{minipage}
  \caption{(Color online) Fixed points and trajectories on stable (bold, red for $a>0$, green for $a<0$) and unstable (thick, dashed) limit cycles in the phase space for the excitable regime (\textbf{A}), the oscillatory regime (\textbf{B}), and the bistable regime (\textbf{C}) for $D=0$ (see Fig. \ref{fig:PhaseDiagram}). 
  Stable nodes (black circles), saddles (white circles), and the velocity field (blue arrows) are depicted. The nullclines for the \mbox{eqs. (\ref{equ.CompleteSystem}) and (\ref{equ.TriggeredFeedback})} are marked by black lines. 
  Red and green colors indicate limit cycles for positive and negative feedback, respectively. Arrows illustrate the corresponding directions. Trajectories reach the corresponding value of $\Delta \omega_{lc}$ at $\phi=2 \pi$, 
  and start with an offset of $2 \pi a/ \tau$ to $\Delta \omega$ after the reset. Here $\Delta \omega^i_{lc}:=\Delta \omega_{lc}(\Delta t_{det}^{(i)})$ according to the eqs. (\ref{equ:solutionForOmega2}) and (\ref{eq:solutionsPeriod}). Parameters: (\textbf{A}) $\omega_{0}=0.8$, $\tau=50$; (\textbf{B}) $\omega_{0}=1.05$, $\tau=50$, $a=\pm 0.5$; (\textbf{C}) $\omega_0=0.85$, $\tau=50$, $a=0.55$.}
 \label{fig:Trajektories}
 \end{figure}
Eq. (\ref{equ.TriggeredFeedback}) describes the dynamics of $\Delta \omega$, evolving on the feedback timescale $\tau$.
Due to the first term, it decays towards zero from any deviation. 
The second term models the feedback and alters $\Delta \omega$ by an amount of $2 \pi a / \tau$ whenever an event occurs ($t = t_{i}$).
This is illustrated in fig. \ref{fig:Trajektories1} for a positive feedback strength $a>0$, showing the time evolution, and in fig. \ref{fig:Trajektories} (center) illustrating the trajectory 
in the $(\phi, \Delta \omega)$-space. Thus, a cycle consists first, of a part, were $\phi$ and $\Delta \omega$ evolve according to the \mbox{eqs. (\ref{equ.CompleteSystem}) and (\ref{equ.TriggeredFeedback})}, respectively.
Secondly, if $\phi$ reaches the threshold, the reset condition eq. (\ref{equ.reset}) is applied. Finally, in the third step, $\Delta \omega$ is altered by an amount of $2 \pi a/\tau$.
Note that putting $a=0$, yields in the stationary case always the situation without feedback.



After some transient behavior, the rate becomes stationary and we define the oscillator's  mean firing rate, which describes the average rate at which events occur
\begin{align}
\label{equ.Omega}
 r=\frac{\langle \dot{\phi}(t) \rangle}{2 \pi}=\langle x(t) \rangle=\frac{1}{\langle \Delta t_i \rangle}.
\end{align}
Here the average is taken over a time interval large compared to the individual IEIs $\Delta t_i = t_{i+1}-t_{i}$, i.e., the time the oscillator needs to reach $\phi=2 \pi$, when started at $\phi=0$.

By averaging eq. (\ref{equ.TriggeredFeedback}), we obtain:
\begin{align}
\label{equ.av}
\tau \Big\langle \frac{d}{dt}\Delta \omega \Big\rangle = -\langle \Delta \omega \rangle + 2 \pi a \langle x(t) \rangle.
\end{align}
In the stationary case, the left hand side should be zero and we obtain
\begin{align}
\begin{array}{rl}
\label{equ:avomegat}
\langle \Delta \omega \rangle =& 2 \pi a r
\end{array}.
\end{align}
Using eq. (\ref{equ:avomegat}) in the averaged eq. (\ref{equ.CompleteSystem}), yields
\begin{align}
\label{equ.OmegaArel}
r = \frac{\omega_0 - \langle \sin[\phi(t)]\rangle}{2 \pi (1-a)}.
\end{align}
Note that $\phi(t)$ is the solution of eq. (\ref{equ.CompleteSystem}) in the presence of feedback.

Interestingly, the limit of $a \nearrow 1$ leads to infinite $r$ if $\omega_0>1$. In this case the unknown numerator is positive, since $\langle \sin[\phi(t)]\rangle \leq 1$. 
Here $\nearrow$ denotes the left-hand limit. For such strong positive feedback, the deterministic decay of $\Delta \omega$ cannot balance the increase of $\Delta \omega$ due to the kicks after 
each event and the assumption of stationarity $\langle \Delta \dot{\omega} \rangle = 0$ does not hold. To study the stationary regime, we therefore concentrate on $a<1$.

\section{Mean interevent interval}
\label{sec:cyclePeriod}
\subsection{Deterministic case}
\begin{figure}[t]
 \centering
   \includegraphics[width=\linewidth]{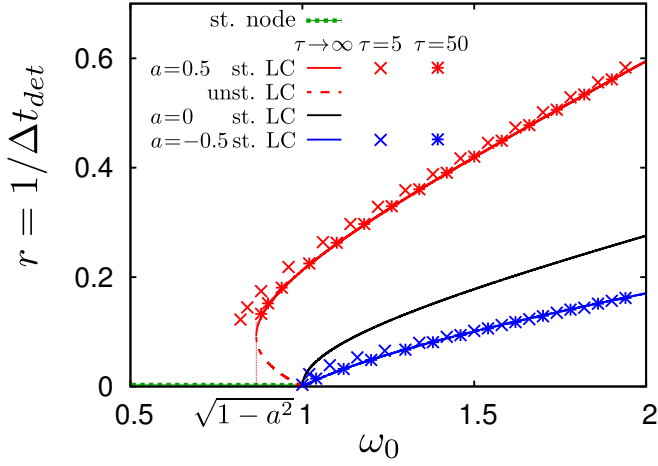}
  \caption{(Color online) Solutions for the steady state firing rate $r$ for $D=0$ in the limit $\tau \rightarrow \infty$ (lines) and results from simulation (points) for $\tau=5$ and $\tau=50$.
For positive feedback $a>0$ (red) two solutions of \mbox{eq. (\ref{eq:solutionsPeriod})} exist and describe oscillations on a stable
(solid, st. LC) and a unstable (dashed, unst. LC) limit cycle, corresponding to the inverse $\Delta t_{det}^{(1),(2)}$, respectively. The solution for $a=0$ (black) is given by eq. (\ref{equ:PeriodKuramoto}). For $a<0$ (blue) only one solution ($\Delta t_{det}^{(1)}$) exist, 
describing oscillations on a stable limit cycle. Independently of $a$ a solution $r=0$ (green, dashed) exists for $\omega_0<1$.}
 \label{fig:RegimesFiring}
 \end{figure}
\begin{figure}[t]
\begin{minipage}[t]{1\linewidth}
 \centering
   \includegraphics[width=\linewidth]{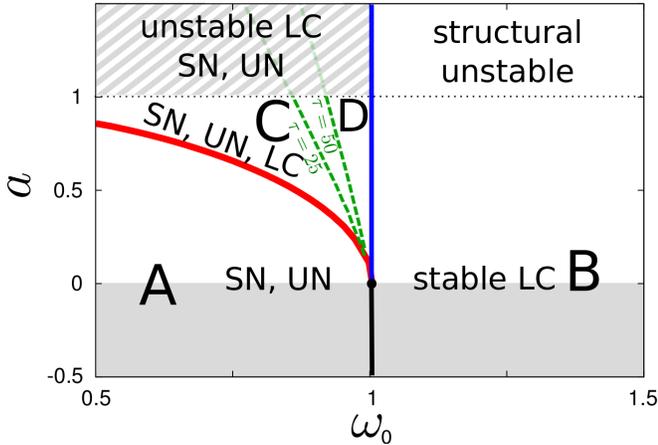}
\end{minipage}
  \caption{(Color online) Dynamical regimes in the $(\omega_0,a)$ parameter space for $\tau \rightarrow \infty$.
  Capital letters name the excitable (\textbf{A}), the oscillatory (\textbf{B}), and the bistable regime (\textbf{C} and \textbf{D}). Topological properties are denoted by SN (stable node), UN (unstable node), and LC (limit cycle).
  Thick lines indicate bifurcations between different regimes, dashed lines bifurcations that occur for finite $\tau$, and the doted line separates the region where trajectories diverge.}
   \label{fig:PhaseDiagram}
 \end{figure}
At first, we concentrate on the deterministic case \mbox{($D=0$)}. Here, after some transient behavior, all IEIs become equal $\Delta t_i=\Delta t_{det}$ for all $i$. 
If no feedback is applied, $\Delta \omega$ will converge to zero and 
the IEIs can be calculated by integrating eq. (\ref{equ.CompleteSystem}), which yields \cite{shinomoto1986phase}
\begin{align}
\label{equ:PeriodKuramoto}
 \Delta t_{det,0}=\frac{2 \pi}{\sqrt{\omega_0^2-1}}.
\end{align} 
Here the index $0$ marks the non-feedback solution for the mean IEI. Note that positive real solutions for $\Delta t_{det,0}$ exist only in the oscillatory regime \mbox{$|\omega_0| > 1$}. 

If, however, feedback is applied ($a \neq 0$), the dynamics becomes more complex. Here the deterministic behavior can be understood by 
evaluating the time-dependent frequency adaptation $\Delta \omega$. Assume, that the system evolves on a LC, and let $\Delta \omega_{lc}$ 
be the value of $\Delta \omega$ just before an event occurs, i.e. 
\begin{align}
\label{equ:LCcondition2}
 \lim \limits_{t \nearrow t_k} \Delta \omega(t) =  & \Delta \omega_{lc}.
\end{align}
After reset, $\Delta \omega$ changes to $\Delta \omega_{lc}+2 \pi a / \tau$, which yields the initial conditions for the next IEI.
Corresponding phase portraits are illustrated
in fig. \ref{fig:Trajektories} (center) for positive (red) and negative feedback (green), respectively.
We can integrate \mbox{eq. (\ref{equ.TriggeredFeedback})} for one IEI, resulting in
\begin{align}
\label{equ:solutionForOmega1}
 \Delta \omega(t)=&(\Delta \omega_{lc}+\frac{2 \pi a}{\tau}) \exp(-\frac{t-t_{k}}{\tau}), & t_k \leq t < t_k+\Delta t_{det}.
\end{align}
Since after one IEI $\Delta \omega$ reaches $\Delta \omega_{lc}$ again, i.e.
\begin{eqnarray}
 \lim \limits_{t \nearrow t_k} \Delta \omega(t) = \lim \limits_{t \nearrow t_k+\Delta t_{det}} \Delta \omega(t) = \Delta \omega_{lc},
\end{eqnarray}
we obtain an explicit expression for $\Delta \omega_{lc}$:
\begin{align}
\label{equ:solutionForOmega2}
 \Delta \omega_{lc}=&\frac{2 \pi a}{\tau [\exp(\frac{\Delta t_{det}}{\tau})-1]}.
\end{align}

In the following, we consider a slow feedback timescale $\tau$, i.e., \mbox{$\tau \gg \Delta t_{det}$}.
In this case, we can expand $\Delta \omega(t)$ [eq. (\ref{equ:solutionForOmega1})] in the small parameter $\Delta t_{det}/\tau$. Using that $t \in [t_k, t_k+\Delta t_{det}[$
and $\Delta \omega_{lc} = 2 \pi a/\Delta t_{det} + { \scriptstyle \mathcal{O}}(\Delta t_{det}/\tau)$ [see eq. (\ref{equ:solutionForOmega2})],
the zeroth order Taylor expansion for $\Delta \omega(t)$ reads
\begin{align}
\label{equ:approximation}
 \Delta \omega(t)=\frac{2 \pi a}{\Delta t_{det}} + { \scriptstyle \mathcal{O}}\left(\frac{\Delta t_{det}}{\tau}\right).
\end{align}
Note that the zeroth order term equals the time-averaged frequency adaptation in eq. (\ref{equ:avomegat}).
Using only the zeroth order in eq. (\ref{equ.CompleteSystem}) for $D=0$, leads to the solvability condition
\begin{align}
 \Delta t_{det} = &\frac{2 \pi}{\sqrt{(\omega_0+ \frac{2 \pi a}{\Delta t_{det}})^2-1}}.
\end{align}



Solving the resulting quadratic equation for $\Delta t_{det}$ yields
\begin{align}
\label{eq:solutionsPeriod}
 \Delta t_{det}^{(1),(2)} \approx \frac{2 \pi}{\omega_0^2-1}(\pm \sqrt{\omega_0^2+(a^2-1)}-a \omega_0), & \ \ \Delta t_{det} \ll \tau.
\end{align} 
By comparison with simulations, we found that positive real solutions $\Delta t_{det}^{(1)}$ correspond to the cycle period of oscillations on a stable limit cycle, whereas 
positive real solutions $\Delta t_{det}^{(2)}$ correspond to the cycle period of oscillations evolving on an unstable limit cycle.

In agreement with eq. (\ref{equ.OmegaArel}), the solution $\Delta t_{det}^{(1)}$ runs to zero (infinite rate) for $a \nearrow 1$ when $\omega_0 \neq 1$.
However, positive solutions $\Delta t_{det}^{(2)}$ also exist for $a \geq 1$, if $\omega_0<1$ (dashed region in fig. \ref{fig:PhaseDiagram}). They describe oscillations on a unstable LC which separates 
the bassin of attraction of the stable node from a regime where the system speeds up to infinite rate.

For $a<1$ we find three qualitatively different regimes. \mbox{Fig. \ref{fig:RegimesFiring}} depicts the resulting firing rates $r=1/\Delta t_{det}^{(i)}$, for \mbox{$i=1,2$}, and 
fig. \ref{fig:PhaseDiagram} illustrates the different regimes in the $(\omega_0,a)$ parameter space.
The corresponding dynamics is illustrated in fig. \ref{fig:Trajektories}.
\begin{itemize}
 \item (\textbf{A}) [Fig. \ref{fig:Trajektories} (top)]: For \mbox{$0 < a < 1$} and \mbox{$\omega_0 < \sqrt{1-a^2}$}, and for $a \leq 0$ and $\omega_0<1$, eq. (\ref{eq:solutionsPeriod}) has no real solution.
		     Here the only stable equilibrium is the SN and only noisy excitations can lead to new events.
		     
 \item (\textbf{B}) [Fig. \ref{fig:Trajektories} (center)]: For \mbox{$a<1$} and \mbox{$\omega_0>1$}, only \mbox{$\Delta t_{det}^{(1)}$} is positive. Here the system possesses a stable LC for 
		    both, negative and positive feedback respectively.
		  
 \item (\textbf{C}) [Fig. \ref{fig:Trajektories} (bottom)]: For \mbox{$0 < a < 1$} and \mbox{$\sqrt{1-a^2} < \omega_0 <1$} eq. (\ref{eq:solutionsPeriod}) has the two positive real solutions $\Delta t_{det}^{(1)}$ and $\Delta t_{det}^{(2)}$.
		    Simulations of trajectories show, that positive solutions of $\Delta t_{det}^{(2)}$ correspond to 
		    slow oscillations on an unstable LC (dashed), which separates the basins of attraction of the stable LC, described by oscillations with period $\Delta t_{det}^{(1)}$ (bold),
		    and the SN (black dot). Here bistability between the SN and the stable LC occurs.
\end{itemize}

These regimes are separated by different bifurcations, indicated by thick lines in fig. \ref{fig:PhaseDiagram}, that can be studied using the positions of the stable $(\phi_{st},\Delta \omega_{st})=(\arcsin[\omega_0],0)$, and unstable node $(\phi_{ust},\Delta \omega_{ust})=(\pi-\arcsin[\omega_0],0)$, and the linearized system of the eqs. (\ref{equ.CompleteSystem}) and (\ref{equ.TriggeredFeedback}) evaluated at fixed $0 < \phi_0 < 2 \pi$ and $\omega_0$
\begin{align}
\label{eq:linearized}
 \left( \begin{matrix}
  \dot{\phi}\\
  \dot{\Delta \omega}
 \end{matrix}\right)=
 \left( 
 \begin{matrix}
  -\cos(\phi_0) & 1\\
  0 & -\frac{1}{\tau}
 \end{matrix}\right) \left( \begin{matrix}
  \phi-\phi_0\\
  \Delta \omega-\Delta \omega_0 
 \end{matrix}\right).
\end{align}
Using the solutions for the mean IEI [eq. (\ref{eq:solutionsPeriod})], one can also study the existence of the limit cycles for $\tau \rightarrow \infty$. The analysis for finite $\tau$ was done by numerical simulations
of trajectories.

For negative feedback (light gray region in fig. \ref{fig:PhaseDiagram}) the regime \textbf{A} transforms into the regime \textbf{B} via saddle-node on invariant circle bifurcation (SNIC) at $(\phi,\Delta \omega)=(\pi/2,0)$
(black line in fig. \ref{fig:PhaseDiagram}). 
A third regime \textbf{C}, and for finite $\tau$ a fourth regime \textbf{D}, exist for $a>0$. 
Here, 
a stable and an unstable LC are born via saddle-node limit
cycle bifurcation at $\sqrt{1-a^2} = \omega_0$ (red line) and the two solutions of eq. (\ref{eq:solutionsPeriod}) coincide. The existence of LCs for finite $\tau$ was verified by simulations for $\tau=25, 50, 100$. For finite $\tau$, the unstable LC approaches the UN and, finally, vanishes via a subcritical Homoclinic orbit bifurcation (SHO) (positive sum of eigenvalues of the Jacobian in eq. (\ref{eq:linearized})) (green, dashed lines), if $\omega_0$ is increased. 
At $\omega_0=1$ the two equilibria annihilate
each other in a saddle-node (off cycle) bifurcation (fold) (blue line). 

In the limit of large $\tau$ the SHO and fold bifurcation occur both at $\omega_0=1$. Here both Eigenvalues of the Jacobian [eq. (\ref{eq:linearized})]
become zero ($-1/\tau \rightarrow 0$), leading to a Bogdanov-Takens bifurcation. However, for finite $\tau$ the $\Delta \omega$ direction is always stable, and the bifurcation at $\omega_0=1$ is of fold ($a > 0$) or SNIC ($a \leq 0$) type. 

We also find, that the rate for finite $\tau$ is higher than in the limit $\tau \rightarrow \infty$ [see fig. \ref{fig:RegimesFiring}] and that the range of bistability extends to smaller values of $\omega_0$.
In the presence of positive feedback, the saddle-node bifurcation of the equilibria at $\omega_0=1$ changes from on cycle (SNIC) to off cycle. 
This affects the system's response to a slowly increasing $\omega_0$ and, therefore, its excitability class \cite{izhikevich2007dynamical}.
At $\omega_0=1$ the SN vanishes and events are produced. Now, the distance to the saddle-node bifurcation is increased by an amount of $2 \pi a / \tau$ after each 
event. For small $\tau$ and large $a$ the system leaves the vicinity of the bifurcation after the first event and produces events at a high rate even for $\omega_0$ close to one.
This leads to class II excitability for a strong positive feedback and small $\tau$.


In the following, we will study the dynamics for $a<1$. We will refer to the regimes as \textit{excitable} (\textbf{A}), \textit{oscillatory} (\textbf{B}) and \textit{bistable} (\textbf{C}, \textbf{D}) according to their properties.
When studying the system in the presence of noise, we concentrate on the \textit{excitable} and the \textit{oscillatory} regime and study how the event-triggered feedback affects the IEI statistics.


\subsection{Finite noise strengths}
In case of finite noise strengths ($D \neq 0$) the mean IEI, of long sequences $\Delta t_i$ ($N \rightarrow \infty$), in the absence of feedback is given by the mean first passage time (FPT) for the system
to reach \mbox{$\phi=2 \pi$} for the first time, when it was started at $\phi=0$.  For this problem, the mean FPT $\langle \Delta t_{i} \rangle$ is given by a well-known integral formula \cite{siegert1951first, anishchenko2007nonlinear} and related to the mean velocity $v$ of a Brownian particle by $\langle \Delta t_{i} \rangle=2 \pi/v$. 
Due to the periodicity of the sinus in eq. (\ref{equ.CompleteSystem}), our system in the absence of feedback is equivalent to overdamped Brownian motion in a tilted periodic potential, for which the mean FPT \cite{Ris84, reimann2002diffusion} is given by
\begin{align}
\label{equ:NoFeedack}
 \frac{1}{r_0}=\langle \Delta t_{i,0} \rangle=\frac{\int \limits_{0}^{2 \pi}dx \ e^{\frac{U_0(x)}{D}}\int \limits_{x-2 \pi}^{x} dy \ e^{-\frac{U_0(y)}{D}}}{D (1-e^{-\frac{2 \pi \omega_0}{D}})}.
\end{align}
Here the index $0$ marks the absence of feedback.
The potential $U_0(\phi)$ is given by \mbox{$U_0(\phi)=-\omega_0 \phi-\cos(\phi)$}. For this potential, eq. (\ref{equ:NoFeedack}) can be written in terms of modified Bessel functions \cite{stratonovich1967}:
\begin{align}
\label{eq:AnalyticResultNoFeedback}
\begin{aligned}
 \langle \Delta t_{i,0} \rangle=\frac{2 \pi^2 |I_{(i \frac{\omega_0}{D})}(\frac{1}{D})|^2}{D \sinh(\frac{\pi \omega_0}{D})}.
 \end{aligned}
\end{align}
Here $I_{n}(y)$ denotes the $n$th modified Bessel function of the first kind. 

In order to account for the feedback, we use the approximation of slow varying $\Delta \omega$ (see above), which
holds in the case of $\langle \Delta t_i \rangle \ll \tau$. For such $\tau$, we can describe the effect
of feedback by substituting $\omega_0 \rightarrow \omega_0+\langle \Delta \omega \rangle$ [compare eq. (\ref{equ:avomegat})] in \mbox{eq. (\ref{equ:NoFeedack})}. 
Applying this 
approximation to $U_0(\phi)$, leads to the extended potential \mbox{$U(\phi)=-(\omega_0 +\langle \Delta \omega \rangle) \phi-\cos(\phi)$}.

Since $\langle \Delta \omega \rangle$ depends on $\langle \Delta t_{i} \rangle$, eq. (\ref{equ:NoFeedack}) 

\clearpage
\onecolumngrid
\begin{figure}[t]
\begin{minipage}{2\linewidth}
\includegraphics[width=\linewidth]{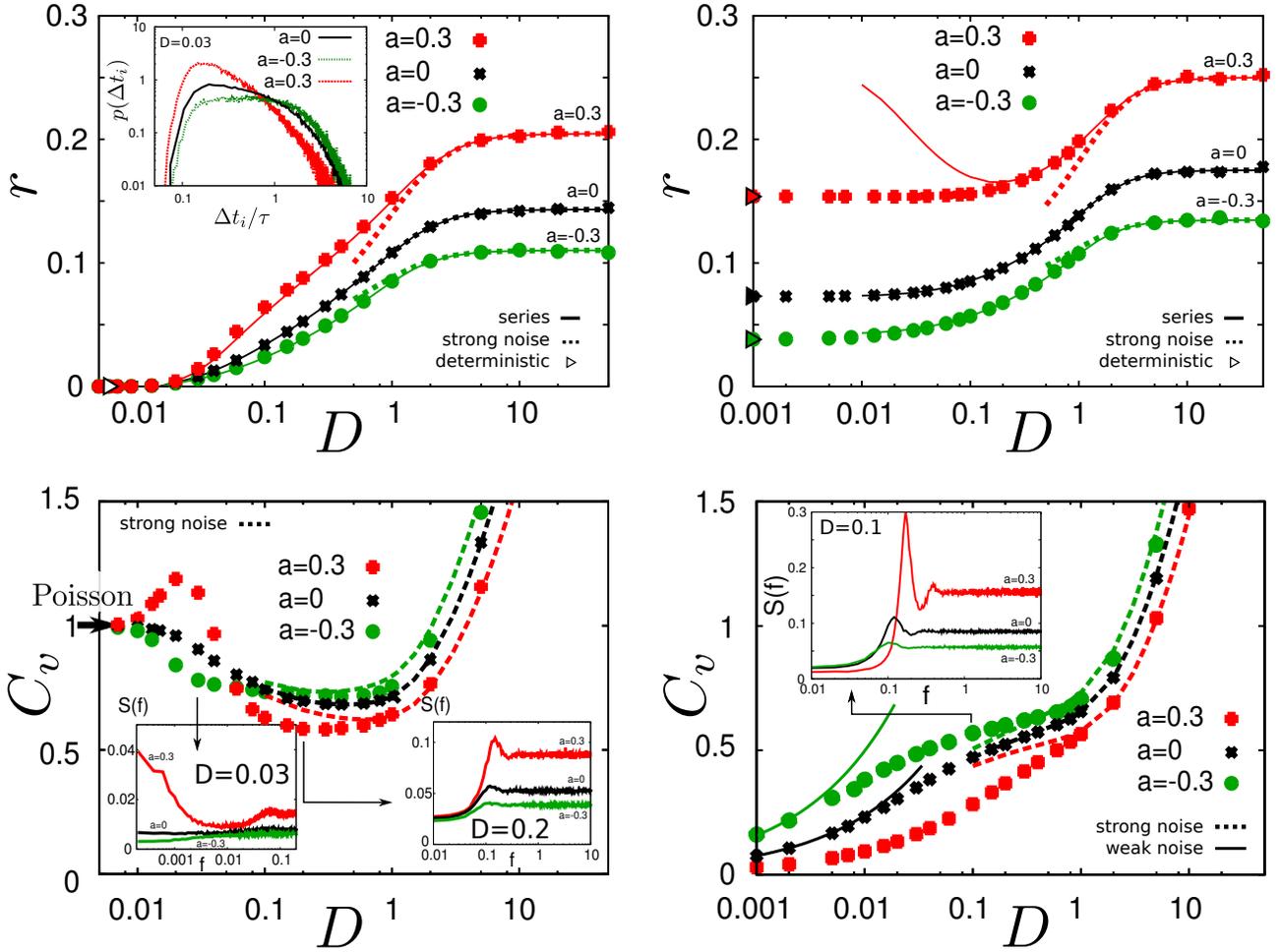}
\caption{(Color online) Firing rate $r$ (top) and CV (bottom) in the excitable regime for $\omega_0=0.9$ (left), and the oscillatory regime for $\omega_0=1.1$ (right), both with $\tau=100$. 
  Insets show the IEI density (top, left) and the power spectra (bottom) for particular noise strengths. 
  Colors denote the particular amount of feedback. Points represent data obtained from simulations. Firing rates (top): Bold lines represent the series approximation \mbox{eq. (\ref{equLowAdLimitMeanISISol})}, 
  dashed lines show the 
  strong noise
  approximation \mbox{eq. (\ref{equLargeNoiseLowAdLimitMeanISI})}  (see details in appendix \ref{sec:VeriWeakNoiseRateFiringRate} for both approximations), and the triangles mark the deterministic firing rates $r=1/\Delta t_{det}^{(1)}$ obtained from eq. (\ref{eq:solutionsPeriod}).
  Firing rates for $D<0.02$ were calculated using the rare-event method presented in Ref. \cite{kromer2013weighted} and are shown in the double logarithmic plot fig. \ref{figWeakNoise} (in appendix \ref{sec:VeriWeakNoiseRateFiringRate})
  together with the weak noise approximation eq. (\ref{equ:kramersratefeedback}),
  CV (bottom): Dashed lines indicate the strong noise approximation [eq. (\ref{equ:R})] and bold lines (right bottom) the weak noise approximation [eq. (\ref{equ:FirstOrderNoiseAndFeedbackCV})]
  (see appendix \ref{sec:VeriWeakNoiseCV} for details).
  In the excitable regime the weak noise limit is given by the Poisson process. Power spectra (bottom, insets) and IEI density (top left, inset) are obtained from simulations.}
 \label{fig:rateCV}
 \end{minipage}
 \end{figure}
\twocolumngrid

%

becomes self-consistent:
\begin{align}
\label{equ:meanIEIFeedack}
\begin{aligned}
 \frac{1}{r}=\langle \Delta t_{i} \rangle=&\frac{\int \limits_{0}^{2 \pi}dx \ e^{\frac{U(x)}{D}} \int \limits_{x-2 \pi}^{x} dy \ e^{\frac{-U(y)}{D}}}{D (1-\exp[{-\frac{2 \pi (\omega_0+\langle \Delta \omega \rangle)}{D}}])}\\
 =&\frac{2 \pi^2 |I_{(i \frac{\omega_0+\langle \Delta \omega \rangle}{D})}(\frac{1}{D})|^2}{D \sinh(\frac{\pi \omega_0+ \pi \langle \Delta \omega \rangle}{D})}, & \langle \Delta t_i \rangle \ll \tau
\end{aligned}
\end{align}
However, for our purpose it is more advantageous to rewrite the integral in eq. (\ref{equ:meanIEIFeedack}) as the series
\begin{align}
\label{exactFPT}
\begin{aligned}
 \langle \Delta t_{i} \rangle=&\frac{2 \pi}{\omega_0+\langle \Delta \omega \rangle} \sum \limits_{k=0}^{\infty}(\frac{1}{2 D})^k\\
  \times  &\sum \limits_{m=0}^{k}\frac{1}{m!(k-m)!}
\frac{I_{|k-2m|}(-\frac{1}{D})}{1+\frac{D^2 (k-2m)^2}{(\omega_0+\langle \Delta \omega \rangle)^2}}, &\langle \Delta t_i \rangle \ll \tau.
\end{aligned}
\end{align}
which can be done after performing some tedious calculations, 
for $\omega_0+\langle \Delta \omega \rangle \neq 0$. 

Assuming a weak feedback $\langle \Delta \omega \rangle / \omega_0 \ll 1$, we can use a Taylor expansion and arrive at the implicit equation
\begin{eqnarray}
\label{equLowAdLimitMeanISI}
\begin{aligned}
\frac{1}{r}=\langle \Delta t_i \rangle \approx \langle \Delta t_{i,0} \rangle-\frac{2 \pi a}{\omega_0}+\frac{4 \pi^2 a}{\omega_0^2 \langle \Delta t_i \rangle} B(D,\omega_0), \notag
\end{aligned}
\\
\begin{aligned}
\langle \Delta \omega \rangle \ll \omega_0, \ \ \langle \Delta t_i \rangle \ll \tau \ \\
\\
\end{aligned}
\end{eqnarray}
for the mean FPT $\langle \Delta t_i \rangle$ in the presence of feedback.

Here $B(D,\omega_0)$ represents the series 
\begin{align}
\begin{aligned}
\label{exactFPTSeriesB}
B(D,\omega_0)=&\sum \limits_{k=1}^{\infty}(\frac{1}{2 D})^k\sum \limits_{m=0}^{k}\frac{1}{m!(k-m)!}\\
&\times \frac{I_{|k-2m|}(-\frac{1}{D})}{1+\frac{D^2}{\omega_0^2}(k-2m)^2}\frac{2(k-2m)^2}{\frac{\omega_0^2}{D^2}+(k-2m)^2}\\
\end{aligned}
\end{align}
If $B(D,\omega_0)$ converges, eq. (\ref{equLowAdLimitMeanISI}) has the only positive solution
\begin{eqnarray}
\label{equLowAdLimitMeanISISol}
\begin{aligned}
\langle \Delta t_i \rangle \approx &\langle \Delta t_{i,0} \rangle-\frac{2 \pi a}{\omega_0}+\frac{4 \pi^2 a}{\omega_0^2 \langle \Delta t_{i,0} \rangle} B(D,\omega_0), \notag
\end{aligned}
\\
\begin{aligned}
\langle \Delta \omega \rangle \ll \omega_0, \ \ \langle \Delta t_i \rangle \ll \tau.
\\
\end{aligned}
\end{eqnarray}
This series is our final result for the mean FPT for large $\tau$ and weak feedback. Its evaluation compared to simulations is illustrated in fig. \ref{fig:rateCV}. 
More details on its evaluation are given in appendix \ref{sec:VeriWeakNoiseRateFiringRate}.

\subsection{Strong noise approximation}
In case of strong noise ($D \gg 1$), the first summand ($k=0$) dominates the series [eq. (\ref{exactFPTSeriesB})] and $\langle \Delta t_i \rangle$ 
can be approximated by \cite{stratonovich1967}
\begin{align}
\label{exactFPTHighNoise}
\frac{1}{r_0}=&\langle \Delta t_{i,0} \rangle \approx \frac{2 \pi}{\omega_0}, &D\gg1,
\end{align}
in the non-feedback case.

Since the function $B(D,\omega)$ is of order $\mathcal{O}(\frac{1}{2 D})$ a 
similar approximation for eq. (\ref{equLowAdLimitMeanISISol}) 
leads to
\begin{align}
\begin{aligned}
\label{equLargeNoiseLowAdLimitMeanISI}
\langle \Delta t_i \rangle \approx  \langle \Delta t_{i,0} \rangle & -\frac{2 \pi a}{\omega_0},\\ 
& \ D\gg1, & \langle \Delta \omega \rangle \ll \omega_0, \ \langle \Delta t_i \rangle \ll \tau.
\end{aligned}
\end{align}
and, in combination with eq. (\ref{exactFPTHighNoise}), to
\begin{align}
\begin{aligned}
\label{equLargeNoiseLowAdLimitMeanISIConst}
\frac{1}{r}=\langle \Delta t_i \rangle = &\frac{2 \pi (1-a)}{\omega_0},\\
& \ \ \ \ \ \ \ D\gg1, & \langle \Delta \omega \rangle \ll \omega_0, \ \langle \Delta t_i \rangle \ll \tau.
\end{aligned}
\end{align} 
Note that this does not depend on the noise strength, like already observed by Stratonovich in the absence of feedback \cite{stratonovich1967}.

\subsection{Weak noise approximation}

In the excitable regime ($\omega_0<1$), the mean IEI in the weak noise limit can be obtained from the Kramers rate theory \cite{hanggi1990reaction}. 
In the absence of feedback, the Kramers rate of generating an event
\begin{align}
\label{equ:kramersrate}
 r_0=& \frac{\sqrt{1-\omega_0^2}}{2 \pi} \ e^{-\frac{\Delta U_{0}}{D}}, & \omega_0<1, \ D \ll 1.
\end{align}
Here \mbox{$\Delta U_{0}=U_{0,max}-U_{0,min}$} denotes the height of the potential barrier.
\mbox{$U_{0,min}=-\sqrt{1-\omega_0^2}-\omega_0 \arcsin(\omega_0)$} and \mbox{$U_{0,max}=-\pi \omega_0+\sqrt{1-\omega_0^2}+\omega_0 \arcsin(\omega_0)$} are the values 
of the potential at the saddle and at the stable node in the absence of feedback, respectively. 

If $\tau$ is large compared to the mean IEI, we can account for the feedback by substituting $\omega_0 \rightarrow \omega_0+\langle \Delta \omega \rangle$ in eq. (\ref{equ:kramersrate}).
In the next step, we assume a weak feedback ($\Delta \omega / \omega_0 \ll 1$) and perform a Taylor expansion. The first order approximation for the potential barrier reads
\begin{align}
\begin{aligned}
\Delta U = &\Delta U_{0}-4 \pi a r \arccos(\omega_0),\\
& \ \omega_0<1, \ D \ll 1, \ \langle \Delta \omega \rangle \ll \omega_0, \ \langle \Delta t_i \rangle \ll \tau.
\end{aligned}
\end{align}
Here $\Delta U$ denotes the barrier of the potential $U$ mentioned above. Consequently, the barrier height becomes rate dependent and reduces for positive feedback 
and increases for negative feedback.

Finally, in the presence of feedback the Kramers rate \mbox{[eq. (\ref{equ:kramersrate})]} reads:
\begin{align}
\begin{aligned}
\label{equ:kramersratefeedback}
 r=&r_0 [1-2 \pi a r_0 \omega_0 (\frac{\omega_0}{1-\omega_0^2}-\frac{2}{D} \arccos[\omega_0])],\\
&\ \ \ \ \ \ \omega_0<1, \ D \ll 1, \ \langle \Delta \omega \rangle \ll \omega_0, \ \langle \Delta t_i \rangle \ll \tau.
\end{aligned}
 \end{align}

In the oscillatory regime we find, using the approach of Ref. \cite{arecchi1980transient}, that the first non-zero correction 
to the deterministic mean IEI \mbox{[eq. (\ref{eq:solutionsPeriod})]} is of order $D^2$. 
\subsection{Results obtained from simulations}
Fig. \ref{fig:rateCV} (top) shows the analytical results in the weak noise limit [eq. (\ref{eq:solutionsPeriod})], for a strong noise [eq. (\ref{equLargeNoiseLowAdLimitMeanISI})], as well
as the series approximation [eq. (\ref{equLowAdLimitMeanISISol})] in the excitable (left) and in the oscillatory (right) regime, respectively. 
A double logarithmic plot of the weak noise regime in the excitable regime is shown in appendix \ref{sec:VeriWeakNoiseRate}. Analytical results are compared to stochastic simulations 
of the model [eqs. (\ref{equ.CompleteSystem}) and (\ref{equ.TriggeredFeedback})].
In the excitable regime, the approximations agree well with the simulations. Here, the strong noise approximation is close to the simulation results for $D>1$, whereas the deterministic firing rate approximates well 
the behavior for $D<0.02$. Here, an even better approximation is given by the correction to the Kramers rate eq. (\ref{equ:kramersratefeedback}) (compare fig. \ref{figWeakNoise}, appendix \ref{sec:VeriWeakNoiseRate}). 
The series approximation can be used for all $D$, however, strong positive feedback $a=0.3$ produces deviations from the theoretical result and small values of $D$ require large computation times. 
In the oscillatory regime we find a similar agreement, except
that the series does not fit the simulations for strong positive feedback in the range of low and intermediate noise strengths. 
Here, the assumption of weak feedback ($\langle \Delta \omega \rangle \ll \omega_0$) does not hold anymore.
Note that since $\langle \Delta \omega \rangle$ depends on the mean IEI, the series approximation leads to better results for low firing rates, i.e., in the excitable regime or for a negative feedback.
In general an increasing noise strength decreases the mean IEI, down to a constant value given by \mbox{eq. (\ref{equLargeNoiseLowAdLimitMeanISIConst})}.
\section{Effect of feedback on output variability}
\label{sec:IEIVariability}
In order to study the variability in a series of IEIs two different measures can be used. The first one is the coefficient of variation (CV)
\begin{eqnarray}
\label{eq:GeneralCV}
 C_v=\frac{\sqrt{\langle (\Delta t_i-\langle \Delta t_i \rangle)^2 \rangle}}{\langle \Delta t_i \rangle},
\end{eqnarray}
in which the standard deviation of the $\Delta t_i$ is compared to its mean. Therefore, $C_v=0$ corresponds to the most regular sequence and, consequently, to the most coherent one, whereas
$C_v=1$ is obtained for a completely random spike train, in which all spikes are independent of each other (Poisson process).

As a second measure of spike train regularity, one can study the power spectrum \cite{pakdaman2001coherence}
\begin{eqnarray}
\label{equ:PowerSpectrum}
 S(f)= \int \limits_{-\infty}^{\infty} dt' \langle x(t)x(t+t') \rangle e^{2 \pi i f t'},
\end{eqnarray}
which measures the spectral components of $x(t)$. In the power spectrum, a narrow peak (possibly accompanied by more peaks at higher harmonics) indicates more coherent sequences of $\Delta t_i$.

In order to calculate the CV of $\Delta t_i$, its mean and its standard deviation $\sqrt{\langle (\Delta t_i-\langle \Delta t_i \rangle)^2 \rangle}$ are needed.
We first calculate the variance $\mathrm{Var}(\Delta t_i)$ of $\Delta t_i$.
In the absence of feedback, $\Delta \omega$ will approach zero and we can apply the formula from Ref. \cite{reimann2002diffusion}, which was
derived for the variance of the FPT density in the case of Brownian motion in a tilted periodic potential
\begin{align}
\begin{aligned}
\label{equ:Variance0}
 \mathrm{Var}(\Delta t_i)=&\frac{2}{D^2 [1-\exp(-\frac{2 \pi \omega_0}{D})]^3}\int \limits_{0}^{2 \pi}dv_1 \ e^{\frac{U_0(v_1)}{D}} \\
 &\times \Big [\int \limits_{v_1-2 \pi}^{v_1} dv_2 \ e^{\frac{-U_0(v_2)}{D}} \Big ]^2 \int \limits_{v_1}^{v_1+2 \pi} dy \ e^{\frac{U_0(y)}{D}}.
 \end{aligned}
\end{align}
Here $U_0(x)$ is the potential used in the previous section.
Applying several simple but tedious steps, similar to those used in the prior section, we end up with a series representation, which, after substituting 
$\omega_0 \rightarrow \omega_0+\langle \Delta \omega \rangle$, and a Taylor expansion with respect to the strength of the feedback ($\langle \Delta \omega \rangle / \omega_0$), yields the first order correction
to the variance
\begin{align}
\begin{aligned}
 &\mathrm{Var}(\Delta t_i) \approx \mathrm{Var}(\Delta t_{i,0})\\
 \times & \Big(1+\frac{2 \pi a}{\langle \Delta t_{i,0} \rangle}[-\frac{3}{\omega_0}+\frac{2 D^2}{\mathrm{Var}(\Delta t_{i,0}) \omega_0^3} C(D,\omega_0)] \Big),\\
 & \ \ \ \ \ \ \ \ \ \ \ \ \ \ \ \ \ \ \ \ \ \ \ \ \ \ \ \ \ \ \ \ \ \ \ \ \ \ \langle \Delta \omega \rangle \ll \omega_0, \ \ \langle \Delta t_i \rangle \ll&\tau.
 \end{aligned}
\end{align}
Here $\mathrm{Var}(\Delta t_{i,0})$ denotes the variance in the absence of feedback ($a=0$) and
$C(D,\omega_0)$ is a infinite series.
\subsubsection{Strong noise approximation}
Fortunately, $C(D,\omega_0)$ vanishes in the strong noise limit $D \rightarrow \infty$.
Therefore, we can derive the analytical approximation for the variance
\begin{align}
\begin{aligned}
\label{equ:variance}
 \mathrm{Var}(\Delta t_{i}) \approx &\mathrm{Var}(\Delta t_{i,0})(1-\frac{2 \pi a}{\langle \Delta t_{i,0} \rangle}\frac{3}{\omega_0}),\\
 & \ \ \ \ \ \ \ \ D \gg 1, \ \langle \Delta \omega \rangle \ll \omega_0, \ \langle \Delta t_i \rangle \ll \tau.
 \end{aligned}
\end{align}
for the strong noise regime.
In this regime, the variance decreases for positive and increases for negative feedback.

Using the eqs. (\ref{equ:variance}) and (\ref{equLargeNoiseLowAdLimitMeanISI}), we obtain the first order correction to the CV
\begin{align}
\begin{aligned}
\label{equ:R}
 C_v \approx &C_{v,0} \left(1-\frac{\pi a}{\omega_0 \langle \Delta t_{i} \rangle} \right),\\
 & \ \ \ \ \  \ \ \ \ \ \ \ \ \ \ \ D \gg 1, \ \langle \Delta \omega \rangle \ll \omega_0, \ \langle \Delta t_i \rangle \ll \tau.
 \end{aligned}
\end{align}
for the strong noise and weak feedback. Here $C_{v,0}$ denotes the CV for $a=0$.
Therefore, positive feedback decreases the CV, whereas negative feedback leads to higher variability in the strong noise regime.
Comparing the strong noise approximation \mbox{[eq. (\ref{equ:R})]} to simulation \mbox{[fig. \ref{fig:rateCV} (bottom)]}, we find that
it fits the numerical results well for \mbox{$D>1$}. 
\subsubsection{Weak noise approximation}
In the weak noise limit, we distinguish between the excitable regime, where the IEI statistics is Poisson-like ($C_v \approx 1$), and the oscillatory regime, 
where the results of Ref. \cite{arecchi1980transient} can be applied.
In the latter case, i.e. for $\omega_0>1$ and in the absence of feedback, the first order approximation for the variance reads
\begin{align}
\label{equ:FirstOrderNoisevariance}
\begin{aligned}
 \mathrm{Var}(\Delta t_{i,0}) \approx & 2 \int \limits_{0}^{2 \pi} d\phi \frac{D}{(\omega_0-\sin[\phi])^3}\\
 \approx & 2 \pi D \frac{1+2 \omega_0^2}{(\omega_0^2-1)^{5/2}}, & \omega_0>1, \ D \ll 1.
\end{aligned}
 \end{align}
Using this in the CV and the weak noise approximation for the mean IEI, yields
\begin{align}
\label{equ:FirstOrderNoiseCV}
 C_{v,0}\approx& \sqrt{\frac{D}{2 \pi}} \sqrt{\frac{1+2 \omega_0^2}{(\omega_0^2-1)^{3 /2}}}, & \omega_0>1, \ D \ll 1. 
\end{align}
Using the substitution \mbox{$\omega_0 \rightarrow \omega_0+\langle \Delta \omega \rangle$}, we can account for the feedback in case of a slow feedback timescale \mbox{$\langle \Delta t_{i} \rangle \ll \tau$}.
By assuming a weak feedback $\langle \Delta \omega \rangle \ll \omega_0$, we obtain the $C_v$ up to first order in $D$:
\begin{align}
\begin{aligned}
\label{equ:FirstOrderNoiseAndFeedbackCV}
 C_{v} \approx & C_{v,0}[1-\frac{\pi a}{\langle \Delta t_{i,0} \rangle}\frac{\omega_0^2(7+2 \omega_0^2)}{(\omega_0^2-1)(1+2\omega_0^2)}],\\
 &  \ \ \ \ \ \ \  \ \ \omega_0 > 1, \ D \ll 1, \ \langle \Delta \omega \rangle \ll \omega_0, \ \langle \Delta t_i \rangle \ll \tau.
 \end{aligned}
\end{align}
Consequently, the CV decreases for positive feedback and increases for negative feedback and hence, qualitatively, the effect of the feedback in the oscillatory regime is similar at weak and strong noise 
\mbox{[cf. eq. (\ref{equ:R})]}.

Figure \ref{fig:rateCV} (left bottom) shows the Poisson limit $C_v=1$. However, for slightly larger $D$ the CV varies strongly with the feedback strength. This variation is due to the dynamics of $\Delta \omega$ (see below) and 
cannot be described by our approach for a slow feedback timescale. 
In the oscillatory regime [fig. \ref{fig:rateCV} (right bottom)] the weak noise approximation 
\mbox{[eq. (\ref{equ:FirstOrderNoiseAndFeedbackCV})]} fits the data well for negative feedback and $D<0.005$. For the positive feedback $a=0.3$, however, the weak-feedback approximation seems to break down and, as a consequence of this \mbox{eq. (\ref{equ:FirstOrderNoiseAndFeedbackCV})} produces slightly negative CVs. 
However, we find that for a weaker feedback with $a=0.15$ the approximation fits the numerical results well (data not shown).
\subsection{Excitable regime}
In excitable systems increasing the noise strength does not necessarily result in higher spike train variability. Instead there exists a minimum 
variability at a finite noise level. This phenomenon is known as coherence resonance (CR) and becomes apparent by a local minimum in the CV or 
by a pronounced peak in the power spectrum attained at an optimal value of the noise intensity. CR can occur in excitable systems due to an interplay of at least two different timescales \cite{pikovsky1997coherence,lindner2002maximizing}, and 
has been observed in the noisy Adler's equation without feedback \cite{qian2000stochastic, lindner2004effects} and experimentally in laser systems \cite{Giacomelli2000, UshWun05}, an electric circuit \cite{PosHan99}, a chemical reaction system \cite{miyakawa2002experimental}, and electrochemical systems \cite{kiss2003experiments, santos2004experimental}.

\begin{figure}[t]
\centering
\begin{minipage}[t]{\linewidth}
\includegraphics[width=\linewidth]{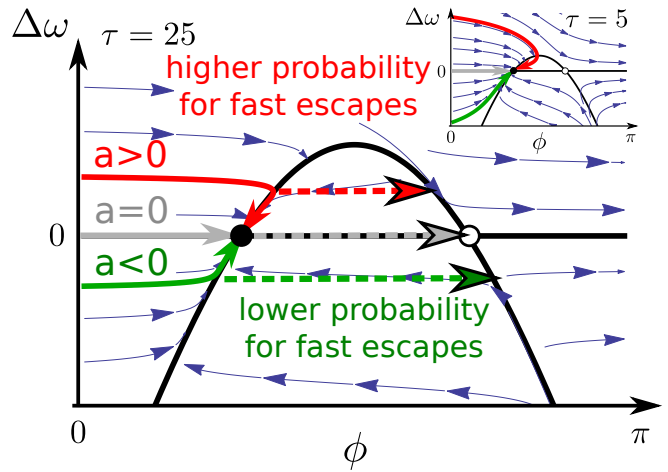}
 \end{minipage}
  \caption{(Color online) Sketch of the dynamics close to the stable node for large $\tau$ and small $\tau$ (inset).
  Black lines represent nullclines, black dots the stable and white dots the unstable nodes, bold colored lines show deterministic trajectories and dashed arrows escapes from the stable branch. 
  Thin blue arrows in the background depict the velocity field $(\dot{\phi}, \Delta \dot{\omega})$.
  Parameters: $\omega_0=0.8$, $a=0.5, 0, -0.5$.
  }
 \label{fig:RdueToOmegaDynamics}
 \end{figure}
Fig. \ref{fig:rateCV} shows the CV (left, bottom) and the power spectrum (left bottom, inset) in the excitable regime for a slow feedback timescale $\tau=100$. 
Here, CR can be observed for intermediate noise strengths, where the $C_v$ possesses a local minimum, already in the absence of feedback ($a=0$). 
In the presence of negative feedback ($a<0$), the $C_v$ slightly increases in those regions but 
reduces for lower noise strengths. Consequently, it increases the region of low $C_v$ towards lower noise strengths. Positive $a$, however, affect $C_v$ in the opposite direction. 
Such feedback improves CR for intermediate noise levels. Note that the CV in our model in the excitable regime is always above $1/\sqrt{3}\approx 0.577$. 
This is similar to a quadratic integrate-and-fire model with noise (but without feedback) and is in marked 
contrast to the range of CV observed in a stochastic leaky integrate-and-fire model \cite{VilLin09}; for differences in signal transmission properties of these models, see  \cite{VilLin09b}.

Interestingly, it also leads to an local maximum of the $C_v$ at a low noise level ($D \approx 0.02-0.03$). Such a maximum indicates \textit{anti-coherence resonance} (ACR) \cite{PhysRevE.66.045105} or \textit{incoherence resonance} \cite{lindner2004effects} and has been observed in models as a consequence of either damped subthreshold oscillations \cite{PhysRevE.66.045105}, or due to a finite refractory period \cite{lindner2002maximizing};
for an experimental verification, in a laser system, see \cite{sergeyev2010coherence}.

For large noise strength the behavior of the $C_v$ can be directly understood from the analytical result eq. (\ref{equ:R}) and is a consequence of the increased or decreased distance to the point ($\omega_0+\langle \Delta \omega \rangle=1$) \cite{lindner2002maximizing}, where the system can pass the 
maximum of the $\phi$-nullcline.

The behavior in the weak noise regime, however, results from the dynamics of $\Delta \omega$, which is illustrated in \mbox{fig. \ref{fig:RdueToOmegaDynamics}} and leads,
in contrast to the oscillatory regime, to a qualitatively different behavior of the CV in the strong and in the weak noise regime, respectively.
For positive feedback, trajectory enter new cycles with positive $\Delta \omega$. Since the $\Delta \omega$-dynamics is usually slower than the $\phi$-dynamics, the system reaches 
the \mbox{$\phi$-nullcline} above the stable node and then relaxes slowly toward the stable fixed point. 
During the relaxation, however, the system can escape the stable node's bassin of attraction much easier than for $\Delta \omega=0$, 
because the distance to the unstable branch and the potential barrier are smaller. This leads to higher probability for small IEIs and a long tail in the FPT density [see fig. \ref{fig:rateCV} (top left, inset)], the latter resulting from the Poisson-like statistics for leaving the SN. 
Since the FPT density shows more probability at times much smaller than the mean IEI, the CV increases \cite{PhysRevE.66.045105}. 
\begin{figure}[t]
\centering
\begin{minipage}[t]{\linewidth}
\includegraphics[width=\linewidth]{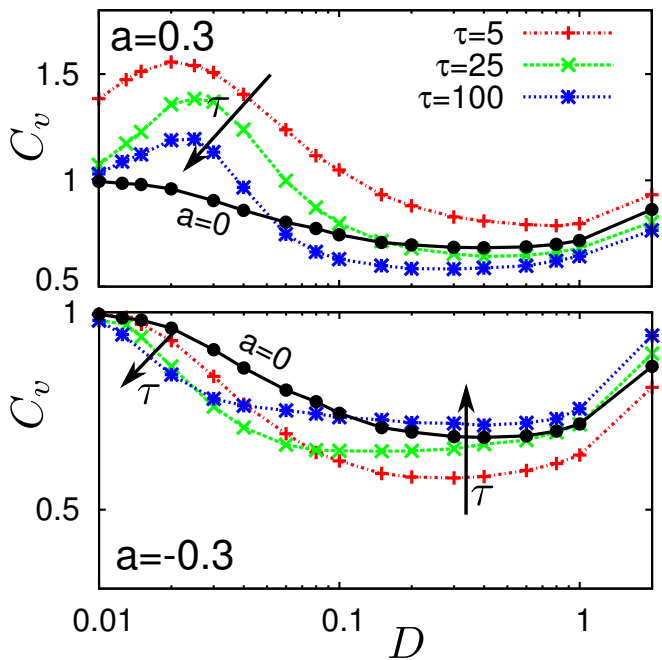}
 \end{minipage}
  \caption{(Color online) Coefficient of variation $C_v$ plotted over the noise strength $D$ for different values of $\tau$ obtained from simulations, for positive $a=0.3$ (top), negative $a=-0.3$ (bottom) and without feedback $a=0$ (black).
  Black arrows indicate the qualitative behavior for an increasing $\tau$.}
 \label{fig:ExcitableCompTau}
 \end{figure}
 
If negative feedback is applied, the trajectories enter new cycles with negative $\Delta \omega$ [see fig. \ref{fig:RdueToOmegaDynamics}]. Once they reach the stable branch the probability for escapes is very low and increases when $\Delta \omega$
relaxes to zero. This reduce the FPT density [fig. \ref{fig:rateCV} (top left, inset)] for times $\Delta t < \tau$ and, consequently, reduces the CV.  

Whether the feedback, finally, enhances or diminishes the CR effect depends on the interplay of both, the CV modulation in the strong noise regime due to the altered
distance to the point $\omega_0+\langle \Delta \omega \rangle=1$, described by \mbox{eq. (\ref{equ:R})}, and the modulation in the weak noise regime, which is dominated by the dynamics of $\Delta \omega$, illustrated in fig. \ref{fig:RdueToOmegaDynamics}. 
\mbox{Fig. \ref{fig:ExcitableCompTau}} shows the CV for different values of $\tau$. Note that a change in the feedback timescale $\tau$ has two effects. On the one hand, it affects the increase of $\Delta \omega$ when an event occurs ($2 \pi a /\tau$) and, on the other hand, it directly alters the timescale separation
between the \mbox{$\phi$-} and the $\Delta \omega$-dynamics. Typical trajectories for a small $\tau$ are depicted in the inset in \mbox{fig. \ref{fig:RdueToOmegaDynamics}}. 
Small $\tau$ enhance the modulation of the CV due to the dynamics of $\Delta \omega$ mentioned above, leading to higher CV in the weak noise (anti-coherence resonance) regime for $a>0$ and to lower CV for $a<0$. 
For negative feedback, the region of low $C_v$ is shifted to higher noise strength when $\tau$ decreases. This occurs due to the larger distance to the point $\omega_0+\langle \Delta \omega \rangle=1$ and has 
been observed in Ref. \cite{lindner2002maximizing}, too.

Analyzing the power spectra, we find very different qualitative behavior in the regions of ACR and CR, respectively. For intermediate noise strength ($D=0.2$) CR occurs \mbox{[see fig. \ref{fig:rateCV}} (left bottom)] and the spectrum
possesses a well pronounced peak. For low frequencies $S(f \rightarrow 0)$, all simulated feedback strengths show quite similar low power. The limit of high frequencies, however, is given by the firing rate $S(f \rightarrow \infty)=r$ and, 
therefore, power increases for positive feedback.
In the region of ACR ($D=0.02-0.03$), the spectrum shows even more interesting behavior. Here, positive feedback leads to more power at low frequencies and increases the power in the peak. 
In between the spectrum possesses a minimum. Consequently, the system operates in two different frequency regimes, possessing bursting behavior. 
Such behavior leads to clusters of small IEIs followed by clusters of large ones.

\subsection{Oscillatory regime}
Fig. \ref{fig:rateCV} (right bottom) shows the CV and the power spectrum (inset) in the oscillatory regime. 
In this regime, the irregularity of spiking increases monotonically  with the noise intensity. Interestingly, positive feedback highly reduces spike train variability
for low and intermediate $D$. Studying the power spectrum, we find that the power at low frequencies is reduced, whereas the peak at $f \approx r$ becomes more pronounced, if positive feedback is applied. 

\section{Feedback-induced correlations}
\label{sec:SerialCorrelations}
The dynamics of $\Delta \omega$ also causes correlations of subsequent IEIs. A measure to quantify correlations of IEIs of lag $n$ is the serial correlation coefficient (SCC) \cite{cox1966statistical}
\begin{eqnarray}
\label{equ:SCC}
 \rho_{n}=\frac{\langle (\Delta t_i - \langle \Delta t_i \rangle) (\Delta t_{i+n}  - \langle \Delta t_i \rangle) \rangle}{\mathrm{Var}(\Delta t_i)}.
\end{eqnarray}
If correlations are positive ($\rho_{n}>0$), longer $\Delta t_i$ are, on average, followed by longer $\Delta t_{i+n}$ (and/or shorter $\Delta t_i$ by shorter $\Delta t_{i+n}$). 
Negative correlations between adjacent intervals ($\rho_1 < 0$) could be caused by an alternation between short and long intervals.
The low frequency limit of the power spectrum is also connected to the SCCs. It holds \cite{cox1966statistical}:
\begin{eqnarray}
\label{equ:LowFreqPower}
 \lim \limits_{f \rightarrow 0}  S(f) =r C_v^2 (1+2 \sum \limits_{k=1}^{\infty} \rho_k).
\end{eqnarray}
Consequently, cumulative IEI correlations can be also studied using the power spectrum.
\subsection{Numerical results in the excitable regime}
\begin{figure}[t]
\centering
 \begin{minipage}[t]{\linewidth}
\centering
  \includegraphics[width=\linewidth]{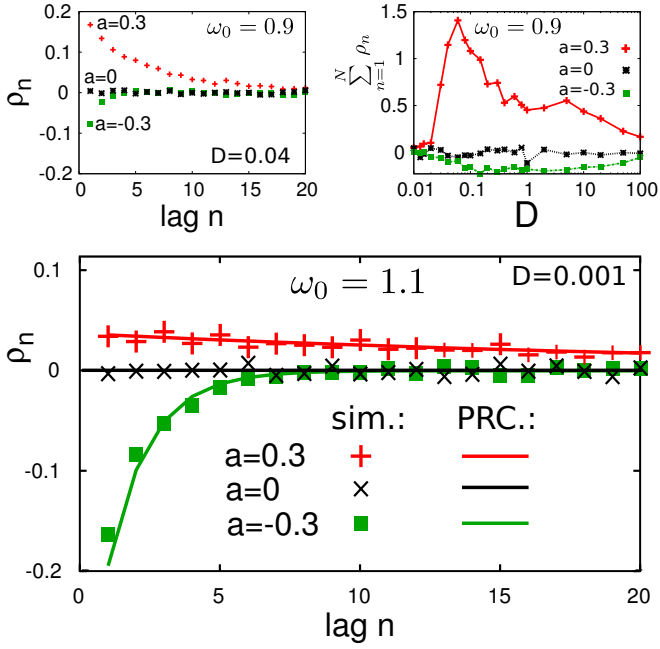}
\end{minipage}
  \caption{(Color online) SCC over several lags (top left) for \mbox{$D=0.04$}, the sum of the first $N=100$ SCCs plotted over noise strength (top right) in the excitable regime ($\omega_0=0.9$) (top),
  and SCC over several lags in the oscillatory regime ($\omega_0=1.1$) (bottom). Excitable regime (top): All results are obtained from simulation. 
  Oscillatory regime (bottom): Numerical results (sim.) are shown 
  together with the analytical approximation eq. (\ref{SCCaprox}).
  Parameter: $\tau=100$.}
 \label{fig:CorOs}
 \end{figure}
 
In the excitable regime, the dynamics of $\Delta \omega$ not only leads to ACR in the weak noise regime, but also causes serial correlations. 
The SCC of several lags is depicted in fig. \ref{fig:CorOs} (top left). It shows strong, slowly-decaying positive and strong, fast-decaying negative correlations for a noise strength
of $D=0.04$, close to the value at which the ACR is observed for positive feedback [compare fig. \ref{fig:rateCV} (left bottom)]. This can be understood by studying the trajectories depicted in fig. \ref{fig:RdueToOmegaDynamics}. 
If positive feedback ($a>0$) is applied, a fast escapes from the stable node, on average, will lead to higher $\Delta \omega$ at $\phi=2 \pi$ [compare fig. \ref{fig:RdueToOmegaDynamics}]. 
Consequently, $\Delta \omega$ will be higher in subsequent IEIs, which increases the probability for fast escapes (small $\Delta t_i$), and, therefore, causes positive 
IEI correlations. However, for negative feedback, the opposite behavior occurs. Here a fast escape (short $\Delta t_i$) leads, on average, to lower $\Delta \omega$
for subsequent cycles and, therefore, further reduce the probability for short $\Delta t_{i+n}$, which leads to negative IEI correlations.

Analyzing the sum of the first $N$ SCCs [fig. \ref{fig:CorOs} (top right)], we find that correlations possess a maximum in the regime of ACR. This can be understood as follows: If $D$ is small, the mean IEI is 
larger than the feedback timescale $\tau$, therefore, perturbations of $\Delta \omega$ are already relaxed when the system can escape the SN, leading to less correlated IEIs. 
However, if $D$ is large, noise dominates the dynamics and leads to less correlations in the sequence, too. Close to the local maximum of the $C_v$, however, we find strong 
positive (for $a>0$) and strong negative (for $a<0$)
serial correlations. 
Combining these findings with eq. (\ref{equ:LowFreqPower}) and using that $C_v$ is of order $1$, we find 
that the increase in the power at low frequencies [see fig. \ref{fig:rateCV} (left bottom, inset)] reflects these correlations.
 
\subsection{Approximation for a slow feedback timescale in the oscillatory regime}

If the system evolves on a limit cycle, the SCC in the weak noise limit can be expressed by a product of the form
\begin{align}
\label{SCCaprox}
 \rho_n=&(\eta V)^{n-1} \rho_1,  & n \geq 1, \ \omega_0>1, \ D \ll 1.
\end{align}
This result was derived for the perfect integrate-and-fire neuron \cite{SchFis10} and for a general integrate-and-fire neuron \cite{schwalger2013patterns}, both 
subjected to an adaptation current (negative feedback), respectively. It can be generalized to positive feedback as long as a limit cycle exists.

The first correlation coefficient $\rho_1$ is given by:
\begin{eqnarray}
\label{equ:rho1}
 \rho_1= -\eta (1-V) \frac{1-\eta^2 V}{1+\eta^2-2 \eta^2 V}.
\end{eqnarray}
Here $\eta$ is determined by the deterministic IEI $\Delta t_{det}$ \mbox{[eq. (\ref{eq:solutionsPeriod})]}
\begin{eqnarray}
 \eta=\exp(-\frac{\Delta t_{det}}{\tau})
\end{eqnarray}
and the term $V$ reads
\begin{eqnarray}
 V=1-\frac{\Delta \omega_{lc}+\frac{2 \pi a}{\tau}}{\tau} \Theta,
\end{eqnarray}
where $\Theta$ is accessible by the phase response curve (PRC) $Z(t)$ \cite{schwalger2013patterns}
\begin{eqnarray}
\label{equ:ThetaPRC1}
 \Theta = - \int \limits_{0}^{\Delta t_{det}} dt \ Z(t) e^{-\frac{t}{\tau}}.
\end{eqnarray}
These formulas have been developed for a perfect \cite{SchFis10} or general multidimensional integrate-and-fire models \cite{schwalger2013patterns} with a spike-triggered linear dynamics for a negative feedback.
We have verified that the approach of Ref. \cite{schwalger2013patterns} also applies to the case of positive feedback as long as a steady state exists, i.e. for $a<1$.

For our system, the PRC can be approximated for a slow feedback timescale ($\Delta t_{det} \ll \tau$) (see Appendix \ref{sec:SerialCorrelationsPRC}).
In this limit, $\Theta$ reads:
\begin{align}
\begin{aligned}
\Theta = &- (1-e^{-\frac{\Delta t_{det}}{\tau}})\frac{\tau}{(\omega_0+\Delta \omega_{lc})} \frac{1+\tau+\tau^2 (\omega_0+\langle \Delta \omega \rangle)^2}{1+\tau^2((\omega_0+\langle \Delta \omega \rangle)^2-1)},\\
 &\ \ \ \ \ \ \ \ \ \ \ \ \ \ \ \ \ \ \ \ \ \ \ \ \ \ \ \ \ \ \ \ \ \ \ \ \ \ \ \ \ \ \ \ \ \ \ \ \ \   \Delta t_{det}\ll \tau.
 \end{aligned}
\end{align}


Here, $\Theta$ is always negative and $\eta$ is close, but smaller than one. 
Consequently, $V$ is larger than one for $a>0$ and smaller than one for $a<0$. This causes $\rho_1$ to have the
same sign as $a$ (compare eq. (\ref{equ:rho1}) for $\eta \lesssim 1$). 
\subsection{Comparison of theory and numerical results in the oscillatory regime}
\subsubsection{Distance to the bifurcation}
\begin{figure}[t]
\begin{minipage}{\linewidth}
  \centering
    \includegraphics[width=\linewidth]{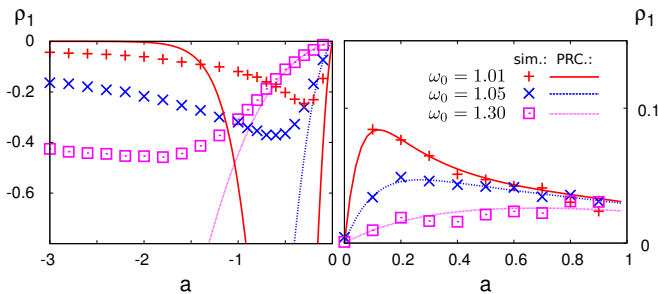}
\end{minipage}
\caption{(Color online) Influence of the distance to the bifurcation on the SCC at lag $1$ for $a<0$ (left) and $a>0$ (right).
The theory eq. (\ref{SCCaprox}) (lines) is compared to simulations for a very low noise level $D=0.001$ (points). 
Close to the bifurcation  ($\omega=1.01$), $\rho_1$ behaves non-monotonically, possessing a local
minimum for negative feedback and a local maximum for positive feedback, respectively. If the distance to the bifurcation is increased ($\omega=1.05$, $\omega=1.3$), the local minimum moves
to stronger negative feedback. For positive feedback, however, the local maximum vanishes in a large distance to the bifurcation. 
Please consider the difference of the total range of $a$ and $\rho_1$ in the two panels. Parameters: $\tau=100$.}
\label{figRho1t100}
\end{figure}
\mbox{Figure \ref{figRho1t100}} shows the analytical results eq. (\ref{equ:rho1}) for $\rho_1$ compared to
those obtained from simulations for different distances to the saddle-node bifurcation at $\omega_0=1$. Interestingly, maximal positive correlations (for $a>0$) become stronger, whereas maximal negative correlations (for $a<0$) 
become weaker by approaching the bifurcation. Note that close to the bifurcation or for strong negative feedback, $\Delta t_{det}$ becomes comparable to $\tau$ \mbox{[see fig. (\ref{fig:RegimesFiring})]}
and the assumption of a slow feedback timescale ($\Delta t_{det} \ll \tau$) does not hold anymore. For this reason the approximation fails quantitatively for large negative values of $a$.

\subsubsection{Non-monotonic behavior}
Another interesting observation can be made in fig. \ref{figRho1t100}: stronger feedback does not necessarily increase $\rho_1$. Instead, the SCC at lag one possesses a minimum for negative
feedback and a maximum when positive feedback is applied. The maximum for positive feedback, however, vanishes if the distance to the bifurcation is increased.

In order to understand how stronger feedback can lead to smaller $\rho_1$, it is helpful to consider the particular trajectories, shown in \mbox{fig. \ref {fig:Trajektories}} (center).
Suppose that the system evolves on the limit cycle and highly negative feedback is applied. In that case, its trajectory looks like the lower one in 
\mbox{fig. \ref {fig:Trajektories}} (center). Such trajectories spend the main part of the IEI close to the stable branch of the $\phi$-nullcline.
The system slowly evolves along the $\phi$-nullcline until $\omega_0+\Delta \omega>1$. Close to the bifurcation ($\omega_0 \gtrsim 0$), however, this
requires $\Delta \omega$ to approach small values.
Consequently, information on perturbations, for instance, due to prior longer (or shorter) IEIs is reduced, which 
decreases $\rho_1$ for strong negative feedback.
In the case of positive or weak negative feedback this effect acts in the opposing direction, since $\Delta \omega$ does not have to increase to pass the maximum of the $\phi$-nullcline. Here, slightly 
higher $\Delta \omega(t_i)$ lead to disproportional shorter IEIs $\Delta t_{i+1}$, whereas initially slightly lower $\Delta \omega(t_i)$ lead to much longer $\Delta t_{i+1}$, if the system is close to the bifurcation point.
Consequently, strong positive correlation between subsequent lags occur. However, for highly positive feedback, the limit cycle is far from the $\phi$-nullcline (see \mbox{fig. \ref{fig:Trajektories}}, center). Here these non-linear effects disappear and $\rho_1$ decreases again.
\subsubsection{Influence of the feedback timescale}
\begin{figure}[t]
\begin{minipage}{1\linewidth}
  \centering
    \includegraphics[width=1\linewidth]{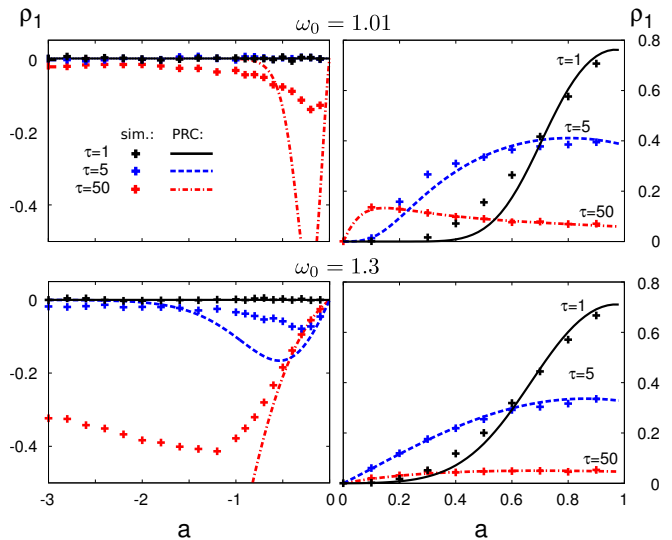}
\end{minipage}
\caption{(Color online) Influence of $\tau$ on the SCC at \mbox{lag $1$} in the weak noise limit for $a<0$ (left) and $a>0$ (right) obtained from
eq. (\ref{SCCaprox}) using eq. (\ref{equ:ThetaPRCFull}) for $\Theta$ (PRC) and from simulations (sim) using $D=0.001$. Close to the bifurcation (top), decreasing $\tau$ 
shifts the maximum of $\rho_1$ to higher $a$. Due to the reduced feedback timescale, subsequent IEIs become more uncorrelated, if weak feedback is applied. However, for $a \rightarrow 1$ the mean IEI runs to zero
and becomes comparable to $\tau$, even with small $\tau$. For negative feedback, small $\tau$ highly reduce $\rho_1$.
}\label{figRho1varyt}
\end{figure}
In fig. \ref{figRho1varyt}, we show  $\rho_1$ as a function of the feedback strength for different $\tau$. Here, interestingly, smaller $\tau$ may lead to stronger correlations for $a>0$,
whereas $\rho_1$ decreases for $a<0$. This occurs due to the increased distance to the $\phi$-nullcline, which leads to smaller IEIs for positive feedback. Since the mean IEI runs to zero for $a \rightarrow 1$,
these correlations are present for strong positive feedback, even for very small $\tau$. 
However, negative feedback leads to larger IEIs, so that perturbations of $\Delta \omega$ cannot survive.
\section{Summary and discussion}
We have studied the effect of event-triggered feedback on the dynamics and output statistics of a noise-driven phase oscillator.

Analytical results for the mean IEIs were derived, which show besides the emergence 
of a bistable regime, that positive feedback leads to a change the bifurcation structure of the system and the excitability class. 
Investigating the influence of the feedback on the output statistics in the excitable regime, we observed that whereas coherence resonance can be observed even without any feedback, only positive feedback leads to 
anti-coherence resonance at low noise strengths.

For both kinds of feedback, we found serial correlations in the sequence of IEIs, which can be approximated analytically in the oscillatory regime for a weak noise and a large timescale separation between the phase and the feedback dynamics, which can be found in cases of spike-triggered feedback due to slow inhibitory currents or slow decaying variations in external ion concentrations in neural systems.
Close to the bifurcation from the excitable to the oscillatory regime, we find a non-monotonic behavior of the correlation between adjacent IEIs and the feedback strength, which indicates that maximal correlations occur
at an optimal feedback strength.
 
Our general approach can be used to understand the role of individual slow processes on the IEI statistics in neurons, excitable lasers, or other pulse-generating systems, that operate close 
to a saddle-node on invariant circle bifurcation (class I excitability), or to identify the source of serial correlations in the IEI sequence. 
Our results illustrate that event-triggered feedback can be used to reduce (or increase) the output variability. This is particularly interesting in information processing systems, in 
which this variability is the limiting factor for a reliable signal transmission. 

\section{Acknowledgments}
This paper was developed within the scope of IRTG 1740/TRP 2011/50151-0, funded by the DFG / FAPESP an by the BMBF (FKZ: 01GQ1001A).

\appendix

\section{Simulation techniques}
\label{sec:simulationtechniques}
All simulations were performed, using the Euler method for the numerical integration of the system \mbox{eqs. (\ref{equ.CompleteSystem}) and (\ref{equ.TriggeredFeedback})}. 
The integration time step was chosen to be $10^{-4}$ for $D<1$ and $10^{-6}$ for larger $D$. After an equilibration time of $100 \tau$, IEIs were recorded up to
an ensemble of $10^{5}$ IEIs. From this series of $\Delta t_i$, the mean firing rate, the CV, the power spectrum, and the SCC was calculated.

In the excitable regime, the firing rate becomes very low, especially for low noise levels. For such weak noise ($D<0.02$), we used the rare event method presented in Ref. \cite{kromer2013weighted}.
Here the parameters, named according to the notation in the reference, read: borders of the simulated area: $L_{\phi}^{-}=-\pi/2$, $L_{\phi}^{-}=-2 \pi$, $L_{\omega}^{-}=-\omega_0$, $L_{\omega}^{+}=1.5$; 
walkers per box: $N=2$; size of a time step $h=0.1$; box size in $\phi$-direction $\Delta \phi = 0.1 \sqrt{2 D h}$; box size in \mbox{$\omega$-direction} $\Delta \tau = 1/(2 \tau)$;.
Simulation were performed for a time $T_{sim}=20000$. After entering the stationary regime, the probability current through absorbing boundary at $\phi=2 \pi$ was recorded and, finally, averaged to
get the mean firing rate.

\section{Details of figure 5}
\label{sec:VeriWeakNoiseRate}
\subsection{Firing rates}
\label{sec:VeriWeakNoiseRateFiringRate}
\textit{The series approximation} was calculated by using eq. (\ref{equLowAdLimitMeanISISol}). For $B(D,\omega_0)$ the terms ($k=1,2,...,500$) were evaluated with high numerical precision.
$\langle \Delta t_{i,0} \rangle$ was obtained from  eq. (\ref{eq:AnalyticResultNoFeedback}). For large $D$ fewer terms are needed to approximate the
firing rate well. However, for $D\approx 0.01$ a few hundred terms are needed and must be calculated with high precision. For even smaller values of $D$ the computation time becomes too large. 
Therefore, the series approximation in fig. \ref{fig:rateCV} (top) is shown for $D\geq 0.01$.

\textit{The strong noise approximation} is given by \mbox{eq. (\ref{equLargeNoiseLowAdLimitMeanISI})}. Here $\langle \Delta t_{i,0} \rangle$ was obtained from  eq. (\ref{eq:AnalyticResultNoFeedback}), too.

\begin{figure}[t]
\begin{minipage}{1\linewidth}
  \centering
    \includegraphics[width=1\linewidth]{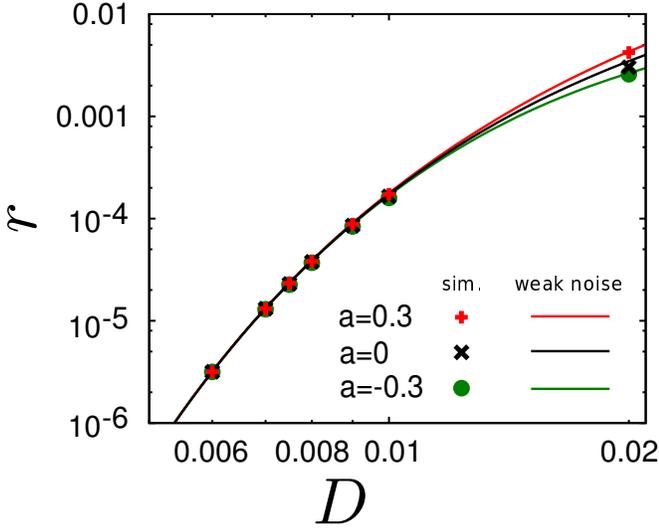}
\end{minipage}
\caption{(Color online) The firing rate in the excitable regime obtained from simulations (points) for a weak noise compared to the analytic approximation eq. (\ref{equ:kramersratefeedback}) (lines). 
Note the double logarithmic scale.
Simulations for $D \leq 0.01$ were performed using the rare-event method presented in Ref. \cite{kromer2013weighted}.
Parameters: $\omega_0=0.9$, $\tau=100$.
}\label{figWeakNoise}
\end{figure}

\textit{The weak noise approximation} eq. (\ref{equ:kramersratefeedback}) was evaluated using eq. (\ref{equ:kramersrate}) for $r_0$ and is illustrated in fig. \ref{figWeakNoise} together with results from simulations. 

\subsection{Coefficient of variation}
\label{sec:VeriWeakNoiseCV}

\textit{The strong noise approximation} was calculated from eq. (\ref{equ:R}), where the eqs. (\ref{eq:AnalyticResultNoFeedback}), (\ref{equ:Variance0}) and (\ref{eq:GeneralCV}) were used for $\langle \Delta t_{i,0} \rangle$ and $C_{v,0}$.

\textit{The weak noise approximation in the oscillatory regime} was obtained from eq. (\ref{equ:FirstOrderNoiseAndFeedbackCV}), where the eqs. (\ref{eq:AnalyticResultNoFeedback}) and (\ref{equ:FirstOrderNoiseCV}) were used for  $\langle \Delta t_{i,0} \rangle$ and $C_{v,0}$.

\section{Calculation of $\Theta$ using the phase response curve}
\label{sec:SerialCorrelationsPRC}
We can calculate $\Theta$, using \mbox{eq. (\ref{equ:ThetaPRC1})}, i.e., by calculating the PRC. In our case the PRC is given by \cite{schwalger2013patterns}:
\begin{align}
\label{equ:PRCcalc}
Z(t)=&Z_{ev}(\Delta t_{det}) \ \exp[-\int \limits_{t}^{\Delta t_{det}} dt' \cos(\phi_{lc}(t'))].
\end{align}
Here $Z_{ev}(\Delta t_{det})=1/(\omega_0+\Delta \omega_{lc})$ is the inverse $\phi$-velocity when an event occurs, if the system evolves on the deterministic limit cycle and $\phi_{lc}(t)$ is the corresponding \mbox{$\phi$-solution}. 

In order to calculate the PRC, we first solve \mbox{eq. (\ref{equ:PRCcalc})} for the non-feedback case and account for feedback by substituting $\omega_0 \rightarrow \omega_0+\langle \Delta \omega \rangle$ afterwards.
In the non-feedback case, $\phi_{lc}(t)$ can be obtained by integrating \mbox{eq. (\ref{equ.noFeedbackCompleteSystem})}, considering $\phi_{lc}(0)=0$ and the smoothness of $\phi_{lc}(t)$ in the interval $t\in [0,\Delta t_{det}]$.
This yields:
\begin{align}
\label{equ:phi}
\phi_{lc}(t) \backsimeq 2 \arctan(\frac{1}{\omega_0}(1-\Omega_0 \tan(\arctan(\frac{1}{\Omega_0})-\frac{\Omega_0}{2} t)).
\end{align}
Here $\Omega_0:=\sqrt{\omega_0^2-1}$ and $\backsimeq$ denotes equality modulo $2 \pi$.
Putting $\phi_{lc}(t)$ into \mbox{eq. (\ref{equ:PRCcalc})}, the PRC in the non-feedback case can be calculated.
After some tedious steps, we get:
\begin{align}
\label{equ:PRC}
Z(t) = Z_{ev}(\Delta t_{det}) (1+\frac{1+ \Omega_0 \sin(\Omega_0 t)- \cos(\Omega_0 t)}{\Omega_0^2}).
\end{align}
The result for $\Theta$ can be obtained from \mbox{eq. (\ref{equ:ThetaPRC1})}. This yields:
\begin{align}
\Theta = - (1-e^{\frac{-\Delta t_{det}}{\tau}}) Z_{ev}(\Delta t_{det}) \tau \frac{1+\tau+\tau^2 \omega_0^2}{1+\tau^2(\omega_0^2-1)}.
\end{align}
Finally, we account for the feedback by substituting \mbox{$\omega_0 \rightarrow \omega_0+\langle \Delta \omega \rangle$}, which yields 
\begin{align}
\label{equ:ThetaPRCFull}
\begin{aligned}
\Theta = &- (1-e^{\frac{-\Delta t_{det}}{\tau}})\frac{\tau}{(\omega_0+\Delta \omega_{lc})} \frac{1+\tau+\tau^2 (\omega_0+\langle \Delta \omega \rangle)^2}{1+\tau^2((\omega_0+\langle \Delta \omega \rangle)^2-1)},\\
& \ \  \ \  \ \  \ \  \ \  \ \  \ \  \ \  \ \  \ \  \ \  \ \  \ \  \ \  \ \  \ \  \ \  \ \  \ \  \ \  \ \  \ \  \ \  \ \  \ \ \tau \gg \Delta t_{det}
\end{aligned}
\end{align}
Expanding this for large $\tau \gg 1$, the zeroth order term reads 
\begin{align}
\label{equ:ThetaPRC}
\Theta \approx &-\frac{\Delta t_{det} (\omega_0+\langle \Delta \omega \rangle)}{(\omega_0+\langle \Delta \omega \rangle)^2-1}, & \ \ \tau \gg \Delta t_{det}.
\end{align}

\renewcommand{\bibname}{references} 
%

\begin{thebibliography}{59}%
\makeatletter
\providecommand \@ifxundefined [1]{%
 \@ifx{#1\undefined}
}%
\providecommand \@ifnum [1]{%
 \ifnum #1\expandafter \@firstoftwo
 \else \expandafter \@secondoftwo
 \fi
}%
\providecommand \@ifx [1]{%
 \ifx #1\expandafter \@firstoftwo
 \else \expandafter \@secondoftwo
 \fi
}%
\providecommand \natexlab [1]{#1}%
\providecommand \enquote  [1]{``#1''}%
\providecommand \bibnamefont  [1]{#1}%
\providecommand \bibfnamefont [1]{#1}%
\providecommand \citenamefont [1]{#1}%
\providecommand \href@noop [0]{\@secondoftwo}%
\providecommand \href [0]{\begingroup \@sanitize@url \@href}%
\providecommand \@href[1]{\@@startlink{#1}\@@href}%
\providecommand \@@href[1]{\endgroup#1\@@endlink}%
\providecommand \@sanitize@url [0]{\catcode `\\12\catcode `\$12\catcode
  `\&12\catcode `\#12\catcode `\^12\catcode `\_12\catcode `\%12\relax}%
\providecommand \@@startlink[1]{}%
\providecommand \@@endlink[0]{}%
\providecommand \url  [0]{\begingroup\@sanitize@url \@url }%
\providecommand \@url [1]{\endgroup\@href {#1}{\urlprefix }}%
\providecommand \urlprefix  [0]{URL }%
\providecommand \Eprint [0]{\href }%
\providecommand \doibase [0]{http://dx.doi.org/}%
\providecommand \selectlanguage [0]{\@gobble}%
\providecommand \bibinfo  [0]{\@secondoftwo}%
\providecommand \bibfield  [0]{\@secondoftwo}%
\providecommand \translation [1]{[#1]}%
\providecommand \BibitemOpen [0]{}%
\providecommand \bibitemStop [0]{}%
\providecommand \bibitemNoStop [0]{.\EOS\space}%
\providecommand \EOS [0]{\spacefactor3000\relax}%
\providecommand \BibitemShut  [1]{\csname bibitem#1\endcsname}%
\let\auto@bib@innerbib\@empty
\bibitem [{\citenamefont {Ermentrout}\ and\ \citenamefont
  {Rinzel}(1984)}]{ermentrout1984beyond}%
  \BibitemOpen
  \bibfield  {author} {\bibinfo {author} {\bibfnamefont {G.~B.}\ \bibnamefont
  {Ermentrout}}\ and\ \bibinfo {author} {\bibfnamefont {J.}~\bibnamefont
  {Rinzel}},\ }\href@noop {} {\bibfield  {journal} {\bibinfo  {journal} {Am. J.
  Physiol.}\ }\textbf {\bibinfo {volume} {246}},\ \bibinfo {pages} {102}
  (\bibinfo {year} {1984})}\BibitemShut {NoStop}%
\bibitem [{\citenamefont {Kuramoto}(1984)}]{kuramoto1984chemical}%
  \BibitemOpen
  \bibfield  {author} {\bibinfo {author} {\bibfnamefont {Y.}~\bibnamefont
  {Kuramoto}},\ }\href@noop {} {\emph {\bibinfo {title} {Chemical Oscillations,
  Waves, and Turbulence}}}\ (\bibinfo  {publisher} {Springer-Verlag, Tokyo},\
  \bibinfo {year} {1984})\BibitemShut {NoStop}%
\bibitem [{\citenamefont {Lindner}\ \emph {et~al.}(2004)\citenamefont
  {Lindner}, \citenamefont {Garc{\i}a-Ojalvo}, \citenamefont {Neiman},\ and\
  \citenamefont {Schimansky-Geier}}]{lindner2004effects}%
  \BibitemOpen
  \bibfield  {author} {\bibinfo {author} {\bibfnamefont {B.}~\bibnamefont
  {Lindner}}, \bibinfo {author} {\bibfnamefont {J.}~\bibnamefont
  {Garc{\i}a-Ojalvo}}, \bibinfo {author} {\bibfnamefont {A.}~\bibnamefont
  {Neiman}}, \ and\ \bibinfo {author} {\bibfnamefont {L.}~\bibnamefont
  {Schimansky-Geier}},\ }\href@noop {} {\bibfield  {journal} {\bibinfo
  {journal} {Phys. Rep.}\ }\textbf {\bibinfo {volume} {392}},\ \bibinfo {pages}
  {321} (\bibinfo {year} {2004})}\BibitemShut {NoStop}%
\bibitem [{\citenamefont {Adler}(1946)}]{adler1946study}%
  \BibitemOpen
  \bibfield  {author} {\bibinfo {author} {\bibfnamefont {R.}~\bibnamefont
  {Adler}},\ }\href@noop {} {\bibfield  {journal} {\bibinfo  {journal} {Proc.
  IRE}\ }\textbf {\bibinfo {volume} {34}},\ \bibinfo {pages} {351} (\bibinfo
  {year} {1946})}\BibitemShut {NoStop}%
\bibitem [{\citenamefont {Giudici}\ \emph {et~al.}(1997)\citenamefont
  {Giudici}, \citenamefont {Green}, \citenamefont {Giacomelli}, \citenamefont
  {Nespolo},\ and\ \citenamefont {Tredicce}}]{giudici1997andronov}%
  \BibitemOpen
  \bibfield  {author} {\bibinfo {author} {\bibfnamefont {M.}~\bibnamefont
  {Giudici}}, \bibinfo {author} {\bibfnamefont {C.}~\bibnamefont {Green}},
  \bibinfo {author} {\bibfnamefont {G.}~\bibnamefont {Giacomelli}}, \bibinfo
  {author} {\bibfnamefont {U.}~\bibnamefont {Nespolo}}, \ and\ \bibinfo
  {author} {\bibfnamefont {J.~R.}\ \bibnamefont {Tredicce}},\ }\href {\doibase
  10.1103/PhysRevE.55.6414} {\bibfield  {journal} {\bibinfo  {journal} {Phys.
  Rev. E}\ }\textbf {\bibinfo {volume} {55}},\ \bibinfo {pages} {6414}
  (\bibinfo {year} {1997})}\BibitemShut {NoStop}%
\bibitem [{\citenamefont {Goulding}\ \emph {et~al.}(2007)\citenamefont
  {Goulding}, \citenamefont {Hegarty}, \citenamefont {Rasskazov}, \citenamefont
  {Melnik}, \citenamefont {Hartnett}, \citenamefont {Greene}, \citenamefont
  {McInerney}, \citenamefont {Rachinskii},\ and\ \citenamefont
  {Huyet}}]{goulding2007excitability}%
  \BibitemOpen
  \bibfield  {author} {\bibinfo {author} {\bibfnamefont {D.}~\bibnamefont
  {Goulding}}, \bibinfo {author} {\bibfnamefont {S.~P.}\ \bibnamefont
  {Hegarty}}, \bibinfo {author} {\bibfnamefont {O.}~\bibnamefont {Rasskazov}},
  \bibinfo {author} {\bibfnamefont {S.}~\bibnamefont {Melnik}}, \bibinfo
  {author} {\bibfnamefont {M.}~\bibnamefont {Hartnett}}, \bibinfo {author}
  {\bibfnamefont {G.}~\bibnamefont {Greene}}, \bibinfo {author} {\bibfnamefont
  {J.~G.}\ \bibnamefont {McInerney}}, \bibinfo {author} {\bibfnamefont
  {D.}~\bibnamefont {Rachinskii}}, \ and\ \bibinfo {author} {\bibfnamefont
  {G.}~\bibnamefont {Huyet}},\ }\href {\doibase 10.1103/PhysRevLett.98.153903}
  {\bibfield  {journal} {\bibinfo  {journal} {Phys. Rev. Lett.}\ }\textbf
  {\bibinfo {volume} {98}},\ \bibinfo {pages} {153903} (\bibinfo {year}
  {2007})}\BibitemShut {NoStop}%
\bibitem [{\citenamefont {Izhikevich}(2007)}]{izhikevich2007dynamical}%
  \BibitemOpen
  \bibfield  {author} {\bibinfo {author} {\bibfnamefont {E.~M.}\ \bibnamefont
  {Izhikevich}},\ }\href@noop {} {\emph {\bibinfo {title} {Dynamical Systems in
  Neuroscience: The Geometry of Excitability and Bursting}}}\ (\bibinfo
  {publisher} {MIT, Cambridge, MA},\ \bibinfo {year} {2007})\BibitemShut
  {NoStop}%
\bibitem [{\citenamefont {Stewart}(1968)}]{stewart1968current}%
  \BibitemOpen
  \bibfield  {author} {\bibinfo {author} {\bibfnamefont {W.}~\bibnamefont
  {Stewart}},\ }\href@noop {} {\bibfield  {journal} {\bibinfo  {journal} {Appl.
  Phys. Lett.}\ }\textbf {\bibinfo {volume} {12}},\ \bibinfo {pages} {277}
  (\bibinfo {year} {1968})}\BibitemShut {NoStop}%
\bibitem [{\citenamefont {Strogatz}(1994)}]{strogatz1994nonlinear}%
  \BibitemOpen
  \bibfield  {author} {\bibinfo {author} {\bibfnamefont {S.~H.}\ \bibnamefont
  {Strogatz}},\ }\href@noop {} {\emph {\bibinfo {title} {Nonlinear Dynamics and
  Chaos: With Applications to Physics, Biology, Chemistry, and Engineering}}}\
  (\bibinfo  {publisher} {Addison-Wesley, Reading, MA},\ \bibinfo {year}
  {1994})\BibitemShut {NoStop}%
\bibitem [{\citenamefont {Aust}\ \emph {et~al.}(2010)\citenamefont {Aust},
  \citenamefont {H{\"o}vel}, \citenamefont {Hizanidis},\ and\ \citenamefont
  {Sch{\"o}ll}}]{aust2010delay}%
  \BibitemOpen
  \bibfield  {author} {\bibinfo {author} {\bibfnamefont {R.}~\bibnamefont
  {Aust}}, \bibinfo {author} {\bibfnamefont {P.}~\bibnamefont {H{\"o}vel}},
  \bibinfo {author} {\bibfnamefont {J.}~\bibnamefont {Hizanidis}}, \ and\
  \bibinfo {author} {\bibfnamefont {E.}~\bibnamefont {Sch{\"o}ll}},\
  }\href@noop {} {\bibfield  {journal} {\bibinfo  {journal} {Eur. J. Phys.
  Special Topics}\ }\textbf {\bibinfo {volume} {187}},\ \bibinfo {pages} {77}
  (\bibinfo {year} {2010})}\BibitemShut {NoStop}%
\bibitem [{\citenamefont {Fourcaud}\ and\ \citenamefont
  {Brunel}(2002)}]{FouBru02}%
  \BibitemOpen
  \bibfield  {author} {\bibinfo {author} {\bibfnamefont {N.}~\bibnamefont
  {Fourcaud}}\ and\ \bibinfo {author} {\bibfnamefont {N.}~\bibnamefont
  {Brunel}},\ }\href@noop {} {\bibfield  {journal} {\bibinfo  {journal} {Neural
  Comput.}\ }\textbf {\bibinfo {volume} {14}},\ \bibinfo {pages} {2057}
  (\bibinfo {year} {2002})}\BibitemShut {NoStop}%
\bibitem [{\citenamefont {Burkitt}(2006)}]{burkitt2006}%
  \BibitemOpen
  \bibfield  {author} {\bibinfo {author} {\bibfnamefont {A.~N.}\ \bibnamefont
  {Burkitt}},\ }\href@noop {} {\bibfield  {journal} {\bibinfo  {journal} {Biol.
  Cybern.}\ }\textbf {\bibinfo {volume} {95}},\ \bibinfo {pages} {1} (\bibinfo
  {year} {2006})}\BibitemShut {NoStop}%
\bibitem [{\citenamefont {Liu}\ and\ \citenamefont {Wang}(2001)}]{LiuWan01}%
  \BibitemOpen
  \bibfield  {author} {\bibinfo {author} {\bibfnamefont {Y.~H.}\ \bibnamefont
  {Liu}}\ and\ \bibinfo {author} {\bibfnamefont {X.~J.}\ \bibnamefont {Wang}},\
  }\href@noop {} {\bibfield  {journal} {\bibinfo  {journal} {J. Comput.
  Neurosci.}\ }\textbf {\bibinfo {volume} {10}},\ \bibinfo {pages} {25}
  (\bibinfo {year} {2001})}\BibitemShut {NoStop}%
\bibitem [{\citenamefont {Benda}\ and\ \citenamefont
  {Herz}(2003)}]{benda2003universal}%
  \BibitemOpen
  \bibfield  {author} {\bibinfo {author} {\bibfnamefont {J.}~\bibnamefont
  {Benda}}\ and\ \bibinfo {author} {\bibfnamefont {A.~V.}\ \bibnamefont
  {Herz}},\ }\href@noop {} {\bibfield  {journal} {\bibinfo  {journal} {Neural
  comp.}\ }\textbf {\bibinfo {volume} {15}},\ \bibinfo {pages} {2523} (\bibinfo
  {year} {2003})}\BibitemShut {NoStop}%
\bibitem [{\citenamefont {Chacron}\ \emph {et~al.}(2005)\citenamefont
  {Chacron}, \citenamefont {Lindner}, \citenamefont {Maler}, \citenamefont
  {Longtin},\ and\ \citenamefont {Bastian}}]{ChaLin05}%
  \BibitemOpen
  \bibfield  {author} {\bibinfo {author} {\bibfnamefont {M.~J.}\ \bibnamefont
  {Chacron}}, \bibinfo {author} {\bibfnamefont {B.}~\bibnamefont {Lindner}},
  \bibinfo {author} {\bibfnamefont {L.}~\bibnamefont {Maler}}, \bibinfo
  {author} {\bibfnamefont {A.}~\bibnamefont {Longtin}}, \ and\ \bibinfo
  {author} {\bibfnamefont {J.}~\bibnamefont {Bastian}},\ }\href@noop {}
  {\bibfield  {journal} {\bibinfo  {journal} {Proc SPIE}\ }\textbf {\bibinfo
  {volume} {5841}},\ \bibinfo {pages} {150} (\bibinfo {year}
  {2005})}\BibitemShut {NoStop}%
\bibitem [{\citenamefont {Schwalger}\ \emph {et~al.}(2010)\citenamefont
  {Schwalger}, \citenamefont {Fisch}, \citenamefont {Benda},\ and\
  \citenamefont {Lindner}}]{SchFis10}%
  \BibitemOpen
  \bibfield  {author} {\bibinfo {author} {\bibfnamefont {T.}~\bibnamefont
  {Schwalger}}, \bibinfo {author} {\bibfnamefont {K.}~\bibnamefont {Fisch}},
  \bibinfo {author} {\bibfnamefont {J.}~\bibnamefont {Benda}}, \ and\ \bibinfo
  {author} {\bibfnamefont {B.}~\bibnamefont {Lindner}},\ }\href@noop {}
  {\bibfield  {journal} {\bibinfo  {journal} {PLoS Comp. Biol.}\ }\textbf
  {\bibinfo {volume} {6}},\ \bibinfo {pages} {e1001026} (\bibinfo {year}
  {2010})}\BibitemShut {NoStop}%
\bibitem [{\citenamefont {Avila-Akerberg}\ and\ \citenamefont
  {Chacron}(2011)}]{AviCha11}%
  \BibitemOpen
  \bibfield  {author} {\bibinfo {author} {\bibfnamefont {O.}~\bibnamefont
  {Avila-Akerberg}}\ and\ \bibinfo {author} {\bibfnamefont {M.~J.}\
  \bibnamefont {Chacron}},\ }\href@noop {} {\bibfield  {journal} {\bibinfo
  {journal} {Exp. Brain Res.}\ }\textbf {\bibinfo {volume} {210}},\ \bibinfo
  {pages} {353} (\bibinfo {year} {2011})}\BibitemShut {NoStop}%
\bibitem [{\citenamefont {Strauss}\ \emph {et~al.}(2008)\citenamefont
  {Strauss}, \citenamefont {Zhou}, \citenamefont {Henning}, \citenamefont
  {Battefeld}, \citenamefont {Wree}, \citenamefont {K{\"o}hling}, \citenamefont
  {Haas}, \citenamefont {Benecke}, \citenamefont {Rolfs},\ and\ \citenamefont
  {Gimsa}}]{strauss2008increasing}%
  \BibitemOpen
  \bibfield  {author} {\bibinfo {author} {\bibfnamefont {U.}~\bibnamefont
  {Strauss}}, \bibinfo {author} {\bibfnamefont {F.}~\bibnamefont {Zhou}},
  \bibinfo {author} {\bibfnamefont {J.}~\bibnamefont {Henning}}, \bibinfo
  {author} {\bibfnamefont {A.}~\bibnamefont {Battefeld}}, \bibinfo {author}
  {\bibfnamefont {A.}~\bibnamefont {Wree}}, \bibinfo {author} {\bibfnamefont
  {R.}~\bibnamefont {K{\"o}hling}}, \bibinfo {author} {\bibfnamefont
  {S.}~\bibnamefont {Haas}}, \bibinfo {author} {\bibfnamefont {R.}~\bibnamefont
  {Benecke}}, \bibinfo {author} {\bibfnamefont {A.}~\bibnamefont {Rolfs}}, \
  and\ \bibinfo {author} {\bibfnamefont {U.}~\bibnamefont {Gimsa}},\
  }\href@noop {} {\bibfield  {journal} {\bibinfo  {journal} {J. Neurophysiol.}\
  }\textbf {\bibinfo {volume} {99}},\ \bibinfo {pages} {2902} (\bibinfo {year}
  {2008})}\BibitemShut {NoStop}%
\bibitem [{\citenamefont {Postnov}\ \emph {et~al.}(2009)\citenamefont
  {Postnov}, \citenamefont {M{\"u}ller}, \citenamefont {Schuppner},\ and\
  \citenamefont {Schimansky-Geier}}]{postnov2009dynamical}%
  \BibitemOpen
  \bibfield  {author} {\bibinfo {author} {\bibfnamefont {D. E.}~\bibnamefont
  {Postnov}}, \bibinfo {author} {\bibfnamefont {F.}~\bibnamefont {M{\"u}ller}},
  \bibinfo {author} {\bibfnamefont {R. B.}~\bibnamefont {Schuppner}}, \ and\
  \bibinfo {author} {\bibfnamefont {L.}~\bibnamefont {Schimansky-Geier}},\
  }\href@noop {} {\bibfield  {journal} {\bibinfo  {journal} {Phys. Rev. E}\
  }\textbf {\bibinfo {volume} {80}},\ \bibinfo {pages} {031921} (\bibinfo
  {year} {2009})}\BibitemShut {NoStop}%
\bibitem [{\citenamefont {Fr{\"o}hlich}\ \emph {et~al.}(2008)\citenamefont
  {Fr{\"o}hlich}, \citenamefont {Bazhenov}, \citenamefont {Iragui-Madoz},\ and\
  \citenamefont {Sejnowski}}]{frohlich2008potassium}%
  \BibitemOpen
  \bibfield  {author} {\bibinfo {author} {\bibfnamefont {F.}~\bibnamefont
  {Fr{\"o}hlich}}, \bibinfo {author} {\bibfnamefont {M.}~\bibnamefont
  {Bazhenov}}, \bibinfo {author} {\bibfnamefont {V.}~\bibnamefont
  {Iragui-Madoz}}, \ and\ \bibinfo {author} {\bibfnamefont {T.~J.}\
  \bibnamefont {Sejnowski}},\ }\href@noop {} {\bibfield  {journal} {\bibinfo
  {journal} {Neuroscientist}\ }\textbf {\bibinfo {volume} {14}},\ \bibinfo
  {pages} {422} (\bibinfo {year} {2008})}\BibitemShut {NoStop}%
\bibitem [{\citenamefont {Fr{\"o}hlich}\ \emph {et~al.}(2006)\citenamefont
  {Fr{\"o}hlich}, \citenamefont {Bazhenov}, \citenamefont {Timofeev},
  \citenamefont {Steriade},\ and\ \citenamefont
  {Sejnowski}}]{frohlich2006slow}%
  \BibitemOpen
  \bibfield  {author} {\bibinfo {author} {\bibfnamefont {F.}~\bibnamefont
  {Fr{\"o}hlich}}, \bibinfo {author} {\bibfnamefont {M.}~\bibnamefont
  {Bazhenov}}, \bibinfo {author} {\bibfnamefont {I.}~\bibnamefont {Timofeev}},
  \bibinfo {author} {\bibfnamefont {M.}~\bibnamefont {Steriade}}, \ and\
  \bibinfo {author} {\bibfnamefont {T.}~\bibnamefont {Sejnowski}},\ }\href@noop
  {} {\bibfield  {journal} {\bibinfo  {journal} {J. Neurosci.}\ }\textbf
  {\bibinfo {volume} {26}},\ \bibinfo {pages} {6153} (\bibinfo {year}
  {2006})}\BibitemShut {NoStop}%
\bibitem [{\citenamefont {Derickson}\ \emph {et~al.}(1990)\citenamefont
  {Derickson}, \citenamefont {Helkey}, \citenamefont {Mar}, \citenamefont
  {Morton},\ and\ \citenamefont {Bowers}}]{derickson1990self}%
  \BibitemOpen
  \bibfield  {author} {\bibinfo {author} {\bibfnamefont {D.~J.}\ \bibnamefont
  {Derickson}}, \bibinfo {author} {\bibfnamefont {R.~J.}\ \bibnamefont
  {Helkey}}, \bibinfo {author} {\bibfnamefont {A.}~\bibnamefont {Mar}},
  \bibinfo {author} {\bibfnamefont {P.~A.}\ \bibnamefont {Morton}}, \ and\
  \bibinfo {author} {\bibfnamefont {J.~E.}\ \bibnamefont {Bowers}},\
  }\href@noop {} {\bibfield  {journal} {\bibinfo  {journal} {Appl. Phys.
  Lett.}\ }\textbf {\bibinfo {volume} {56}},\ \bibinfo {pages} {7} (\bibinfo
  {year} {1990})}\BibitemShut {NoStop}%
\bibitem [{\citenamefont {Schwalger}\ \emph {et~al.}(2012)\citenamefont
  {Schwalger}, \citenamefont {Tiana-Alsina}, \citenamefont {Torrent},
  \citenamefont {Garcia-Ojalvo},\ and\ \citenamefont
  {Lindner}}]{schwalger2012interspike}%
  \BibitemOpen
  \bibfield  {author} {\bibinfo {author} {\bibfnamefont {T.}~\bibnamefont
  {Schwalger}}, \bibinfo {author} {\bibfnamefont {J.}~\bibnamefont
  {Tiana-Alsina}}, \bibinfo {author} {\bibfnamefont {M.}~\bibnamefont
  {Torrent}}, \bibinfo {author} {\bibfnamefont {J.}~\bibnamefont
  {Garcia-Ojalvo}}, \ and\ \bibinfo {author} {\bibfnamefont {B.}~\bibnamefont
  {Lindner}},\ }\href@noop {} {\bibfield  {journal} {\bibinfo  {journal} {Eur.
  Phys. Lett.}\ }\textbf {\bibinfo {volume} {99}},\ \bibinfo {pages} {10004}
  (\bibinfo {year} {2012})}\BibitemShut {NoStop}%
\bibitem [{\citenamefont {Ozbudak}\ \emph {et~al.}(2004)\citenamefont
  {Ozbudak}, \citenamefont {Thattai}, \citenamefont {Lim}, \citenamefont
  {Shraiman},\ and\ \citenamefont
  {Van~Oudenaarden}}]{ozbudak2004multistability}%
  \BibitemOpen
  \bibfield  {author} {\bibinfo {author} {\bibfnamefont {E.~M.}\ \bibnamefont
  {Ozbudak}}, \bibinfo {author} {\bibfnamefont {M.}~\bibnamefont {Thattai}},
  \bibinfo {author} {\bibfnamefont {H.~N.}\ \bibnamefont {Lim}}, \bibinfo
  {author} {\bibfnamefont {B.~I.}\ \bibnamefont {Shraiman}}, \ and\ \bibinfo
  {author} {\bibfnamefont {A.}~\bibnamefont {Van~Oudenaarden}},\ }\href@noop {}
  {\bibfield  {journal} {\bibinfo  {journal} {Nature}\ }\textbf {\bibinfo
  {volume} {427}},\ \bibinfo {pages} {737} (\bibinfo {year}
  {2004})}\BibitemShut {NoStop}%
\bibitem [{\citenamefont {Choi}\ \emph {et~al.}(2008)\citenamefont {Choi},
  \citenamefont {Cai}, \citenamefont {Frieda},\ and\ \citenamefont
  {Xie}}]{choi2008stochastic}%
  \BibitemOpen
  \bibfield  {author} {\bibinfo {author} {\bibfnamefont {P.~J.}\ \bibnamefont
  {Choi}}, \bibinfo {author} {\bibfnamefont {L.}~\bibnamefont {Cai}}, \bibinfo
  {author} {\bibfnamefont {K.}~\bibnamefont {Frieda}}, \ and\ \bibinfo {author}
  {\bibfnamefont {X.~S.}\ \bibnamefont {Xie}},\ }\href@noop {} {\bibfield
  {journal} {\bibinfo  {journal} {Science}\ }\textbf {\bibinfo {volume}
  {322}},\ \bibinfo {pages} {442} (\bibinfo {year} {2008})}\BibitemShut
  {NoStop}%
\bibitem [{\citenamefont {Ermentrout}(1998)}]{ermentrout1998linearization}%
  \BibitemOpen
  \bibfield  {author} {\bibinfo {author} {\bibfnamefont {B.}~\bibnamefont
  {Ermentrout}},\ }\href@noop {} {\bibfield  {journal} {\bibinfo  {journal}
  {Neural Comp.}\ }\textbf {\bibinfo {volume} {10}},\ \bibinfo {pages} {1721}
  (\bibinfo {year} {1998})}\BibitemShut {NoStop}%
\bibitem [{\citenamefont
  {Urdapilleta}(2011{\natexlab{a}})}]{urdapilleta2011survival}%
  \BibitemOpen
  \bibfield  {author} {\bibinfo {author} {\bibfnamefont {E.}~\bibnamefont
  {Urdapilleta}},\ }\href@noop {} {\bibfield  {journal} {\bibinfo  {journal}
  {Phys. Rev. E}\ }\textbf {\bibinfo {volume} {83}},\ \bibinfo {pages} {021102}
  (\bibinfo {year} {2011}{\natexlab{a}})}\BibitemShut {NoStop}%
\bibitem [{\citenamefont {Schwalger}\ \emph {et~al.}(2013)\citenamefont
  {Schwalger}, \citenamefont {Miklody},\ and\ \citenamefont
  {Lindner}}]{schwalger2013leak}%
  \BibitemOpen
  \bibfield  {author} {\bibinfo {author} {\bibfnamefont {T.}~\bibnamefont
  {Schwalger}}, \bibinfo {author} {\bibfnamefont {D.}~\bibnamefont {Miklody}},
  \ and\ \bibinfo {author} {\bibfnamefont {B.}~\bibnamefont {Lindner}},\
  }\href@noop {} {\bibfield  {journal} {\bibinfo  {journal} {Eur. Phys. J.
  Special Topics}\ }\textbf {\bibinfo {volume} {222}},\ \bibinfo {pages} {2655}
  (\bibinfo {year} {2013})}\BibitemShut {NoStop}%
\bibitem [{\citenamefont {Urdapilleta}(2011{\natexlab{b}})}]{Urd11}%
  \BibitemOpen
  \bibfield  {author} {\bibinfo {author} {\bibfnamefont {E.}~\bibnamefont
  {Urdapilleta}},\ }\href@noop {} {\bibfield  {journal} {\bibinfo  {journal}
  {Phys. Rev. E.}\ }\textbf {\bibinfo {volume} {84}},\ \bibinfo {pages}
  {041904} (\bibinfo {year} {2011}{\natexlab{b}})}\BibitemShut {NoStop}%
\bibitem [{\citenamefont {Schwalger}\ and\ \citenamefont
  {Lindner}(2013)}]{schwalger2013patterns}%
  \BibitemOpen
  \bibfield  {author} {\bibinfo {author} {\bibfnamefont {T.}~\bibnamefont
  {Schwalger}}\ and\ \bibinfo {author} {\bibfnamefont {B.}~\bibnamefont
  {Lindner}},\ }\href@noop {} {\bibfield  {journal} {\bibinfo  {journal}
  {Front. Comp. Neurosci.}\ }\textbf {\bibinfo {volume} {7}},\ \bibinfo {pages}
  {164} (\bibinfo {year} {2013})}\BibitemShut {NoStop}%
\bibitem [{\citenamefont {Chacron}\ \emph {et~al.}(2004)\citenamefont
  {Chacron}, \citenamefont {Lindner},\ and\ \citenamefont
  {Longtin}}]{chacron2004noise}%
  \BibitemOpen
  \bibfield  {author} {\bibinfo {author} {\bibfnamefont {M.~J.}\ \bibnamefont
  {Chacron}}, \bibinfo {author} {\bibfnamefont {B.}~\bibnamefont {Lindner}}, \
  and\ \bibinfo {author} {\bibfnamefont {A.}~\bibnamefont {Longtin}},\
  }\href@noop {} {\bibfield  {journal} {\bibinfo  {journal} {Phys. Rev. Lett.}\
  }\textbf {\bibinfo {volume} {92}},\ \bibinfo {pages} {080601} (\bibinfo
  {year} {2004})}\BibitemShut {NoStop}%
\bibitem [{\citenamefont {Chacron}\ \emph {et~al.}(2007)\citenamefont
  {Chacron}, \citenamefont {Lindner},\ and\ \citenamefont
  {Longtin}}]{ChaLin07}%
  \BibitemOpen
  \bibfield  {author} {\bibinfo {author} {\bibfnamefont {M.}~\bibnamefont
  {Chacron}}, \bibinfo {author} {\bibfnamefont {B.}~\bibnamefont {Lindner}}, \
  and\ \bibinfo {author} {\bibfnamefont {A.}~\bibnamefont {Longtin}},\
  }\href@noop {} {\bibfield  {journal} {\bibinfo  {journal} {J. Comput.
  Neuroscie.}\ }\textbf {\bibinfo {volume} {23}},\ \bibinfo {pages} {301}
  (\bibinfo {year} {2007})}\BibitemShut {NoStop}%
\bibitem [{\citenamefont {Chacron}\ \emph {et~al.}(2000)\citenamefont
  {Chacron}, \citenamefont {Longtin}, \citenamefont {St-Hilaire},\ and\
  \citenamefont {Maler}}]{ChaLon00}%
  \BibitemOpen
  \bibfield  {author} {\bibinfo {author} {\bibfnamefont {M.~J.}\ \bibnamefont
  {Chacron}}, \bibinfo {author} {\bibfnamefont {A.}~\bibnamefont {Longtin}},
  \bibinfo {author} {\bibfnamefont {M.}~\bibnamefont {St-Hilaire}}, \ and\
  \bibinfo {author} {\bibfnamefont {L.}~\bibnamefont {Maler}},\ }\href@noop {}
  {\bibfield  {journal} {\bibinfo  {journal} {Phys. Rev. Lett.}\ }\textbf
  {\bibinfo {volume} {85}},\ \bibinfo {pages} {1576} (\bibinfo {year}
  {2000})}\BibitemShut {NoStop}%
\bibitem [{\citenamefont {Prescott}\ and\ \citenamefont
  {Sejnowski}(2008)}]{PreSej08}%
  \BibitemOpen
  \bibfield  {author} {\bibinfo {author} {\bibfnamefont {S.~A.}\ \bibnamefont
  {Prescott}}\ and\ \bibinfo {author} {\bibfnamefont {T.~J.}\ \bibnamefont
  {Sejnowski}},\ }\href@noop {} {\bibfield  {journal} {\bibinfo  {journal} {J.
  Neurosci.}\ }\textbf {\bibinfo {volume} {28}},\ \bibinfo {pages} {13649}
  (\bibinfo {year} {2008})}\BibitemShut {NoStop}%
\bibitem [{\citenamefont {Farkhooi}\ \emph {et~al.}(2009)\citenamefont
  {Farkhooi}, \citenamefont {Strube-Bloss},\ and\ \citenamefont
  {Nawrot}}]{farkhooi2009serial}%
  \BibitemOpen
  \bibfield  {author} {\bibinfo {author} {\bibfnamefont {F.}~\bibnamefont
  {Farkhooi}}, \bibinfo {author} {\bibfnamefont {M.~F.}\ \bibnamefont
  {Strube-Bloss}}, \ and\ \bibinfo {author} {\bibfnamefont {M.~P.}\
  \bibnamefont {Nawrot}},\ }\href {\doibase 10.1103/PhysRevE.79.021905}
  {\bibfield  {journal} {\bibinfo  {journal} {Phys. Rev. E}\ }\textbf {\bibinfo
  {volume} {79}},\ \bibinfo {pages} {021905} (\bibinfo {year}
  {2009})}\BibitemShut {NoStop}%
\bibitem [{\citenamefont {Shinomoto}\ and\ \citenamefont
  {Kuramoto}(1986)}]{shinomoto1986phase}%
  \BibitemOpen
  \bibfield  {author} {\bibinfo {author} {\bibfnamefont {S.}~\bibnamefont
  {Shinomoto}}\ and\ \bibinfo {author} {\bibfnamefont {Y.}~\bibnamefont
  {Kuramoto}},\ }\href@noop {} {\bibfield  {journal} {\bibinfo  {journal}
  {Prog. Theor. Phys.}\ }\textbf {\bibinfo {volume} {75}},\ \bibinfo {pages}
  {1105} (\bibinfo {year} {1986})}\BibitemShut {NoStop}%
\bibitem [{\citenamefont {Siegert}(1951)}]{siegert1951first}%
  \BibitemOpen
  \bibfield  {author} {\bibinfo {author} {\bibfnamefont {A.~J.}\ \bibnamefont
  {Siegert}},\ }\href@noop {} {\bibfield  {journal} {\bibinfo  {journal} {Phys.
  Rev.}\ }\textbf {\bibinfo {volume} {81}},\ \bibinfo {pages} {617} (\bibinfo
  {year} {1951})}\BibitemShut {NoStop}%
\bibitem [{\citenamefont {Anishchenko}\ \emph {et~al.}(2007)\citenamefont
  {Anishchenko}, \citenamefont {Astakhov}, \citenamefont {Neiman},
  \citenamefont {Vadivasova},\ and\ \citenamefont
  {Schimansky-Geier}}]{anishchenko2007nonlinear}%
  \BibitemOpen
  \bibfield  {author} {\bibinfo {author} {\bibfnamefont {V.~S.}\ \bibnamefont
  {Anishchenko}}, \bibinfo {author} {\bibfnamefont {V.}~\bibnamefont
  {Astakhov}}, \bibinfo {author} {\bibfnamefont {A.}~\bibnamefont {Neiman}},
  \bibinfo {author} {\bibfnamefont {T.}~\bibnamefont {Vadivasova}}, \ and\
  \bibinfo {author} {\bibfnamefont {L.}~\bibnamefont {Schimansky-Geier}},\
  }\href@noop {} {\emph {\bibinfo {title} {Nonlinear Dynamics of Chaotic and
  Stochastic Systems: Tutorial and Modern Developments}}}\ (\bibinfo
  {publisher} {Springer, Berlin Heidelberg},\ \bibinfo {year}
  {2007})\BibitemShut {NoStop}%
\bibitem [{\citenamefont {Risken}(1984)}]{Ris84}%
  \BibitemOpen
  \bibfield  {author} {\bibinfo {author} {\bibfnamefont {H.}~\bibnamefont
  {Risken}},\ }\href@noop {} {\emph {\bibinfo {title} {The Fokker-Planck
  Equation}}}\ (\bibinfo  {publisher} {Springer},\ \bibinfo {address}
  {Berlin},\ \bibinfo {year} {1984})\BibitemShut {NoStop}%
\bibitem [{\citenamefont {Reimann}\ \emph {et~al.}(2002)\citenamefont
  {Reimann}, \citenamefont {Van~den Broeck}, \citenamefont {Linke},
  \citenamefont {H{\"a}nggi}, \citenamefont {Rubi},\ and\ \citenamefont
  {P{\'e}rez-Madrid}}]{reimann2002diffusion}%
  \BibitemOpen
  \bibfield  {author} {\bibinfo {author} {\bibfnamefont {P.}~\bibnamefont
  {Reimann}}, \bibinfo {author} {\bibfnamefont {C.}~\bibnamefont {Van~den
  Broeck}}, \bibinfo {author} {\bibfnamefont {H.}~\bibnamefont {Linke}},
  \bibinfo {author} {\bibfnamefont {P.}~\bibnamefont {H{\"a}nggi}}, \bibinfo
  {author} {\bibfnamefont {J. M.}~\bibnamefont {Rubi}}, \ and\ \bibinfo {author}
  {\bibfnamefont {A.}~\bibnamefont {P{\'e}rez-Madrid}},\ }\href@noop {}
  {\bibfield  {journal} {\bibinfo  {journal} {Phys. Rev. E}\ }\textbf {\bibinfo
  {volume} {65}},\ \bibinfo {pages} {031104} (\bibinfo {year}
  {2002})}\BibitemShut {NoStop}%
\bibitem [{\citenamefont {Stratonovich}(1967)}]{stratonovich1967}%
  \BibitemOpen
  \bibfield  {author} {\bibinfo {author} {\bibfnamefont {R.~L.}\ \bibnamefont
  {Stratonovich}},\ }\href@noop {} {\emph {\bibinfo {title} {Topics in the
  Theory of Random Noise}}},\ Vol.~\bibinfo {volume} {2}\ (\bibinfo
  {publisher} {Gordon and Breach, New York},\ \bibinfo {year}
  {1967})\BibitemShut {NoStop}%
\bibitem [{\citenamefont {Kromer}\ \emph {et~al.}(2013)\citenamefont {Kromer},
  \citenamefont {Schimansky-Geier},\ and\ \citenamefont
  {Toral}}]{kromer2013weighted}%
  \BibitemOpen
  \bibfield  {author} {\bibinfo {author} {\bibfnamefont {J.~A.}\ \bibnamefont
  {Kromer}}, \bibinfo {author} {\bibfnamefont {L.}~\bibnamefont
  {Schimansky-Geier}}, \ and\ \bibinfo {author} {\bibfnamefont
  {R.}~\bibnamefont {Toral}},\ }\href {\doibase 10.1103/PhysRevE.87.063311}
  {\bibfield  {journal} {\bibinfo  {journal} {Phys. Rev. E}\ }\textbf {\bibinfo
  {volume} {87}},\ \bibinfo {pages} {063311} (\bibinfo {year}
  {2013})}\BibitemShut {NoStop}%
\bibitem [{\citenamefont {H{\"a}nggi}\ \emph {et~al.}(1990)\citenamefont
  {H{\"a}nggi}, \citenamefont {Talkner},\ and\ \citenamefont
  {Borkovec}}]{hanggi1990reaction}%
  \BibitemOpen
  \bibfield  {author} {\bibinfo {author} {\bibfnamefont {P.}~\bibnamefont
  {H{\"a}nggi}}, \bibinfo {author} {\bibfnamefont {P.}~\bibnamefont {Talkner}},
  \ and\ \bibinfo {author} {\bibfnamefont {M.}~\bibnamefont {Borkovec}},\
  }\href@noop {} {\bibfield  {journal} {\bibinfo  {journal} {Rev. Mod. Phys.}\
  }\textbf {\bibinfo {volume} {62}},\ \bibinfo {pages} {251} (\bibinfo {year}
  {1990})}\BibitemShut {NoStop}%
\bibitem [{\citenamefont {Arecchi}\ and\ \citenamefont
  {Politi}(1980)}]{arecchi1980transient}%
  \BibitemOpen
  \bibfield  {author} {\bibinfo {author} {\bibfnamefont {F. T.}~\bibnamefont
  {Arecchi}}\ and\ \bibinfo {author} {\bibfnamefont {A.}~\bibnamefont
  {Politi}},\ }\href {\doibase 10.1103/PhysRevLett.45.1219} {\bibfield
  {journal} {\bibinfo  {journal} {Phys. Rev. Lett.}\ }\textbf {\bibinfo
  {volume} {45}},\ \bibinfo {pages} {1219} (\bibinfo {year}
  {1980})}\BibitemShut {NoStop}%
\bibitem [{\citenamefont {Pakdaman}\ \emph {et~al.}(2001)\citenamefont
  {Pakdaman}, \citenamefont {Tanabe},\ and\ \citenamefont
  {Shimokawa}}]{pakdaman2001coherence}%
  \BibitemOpen
  \bibfield  {author} {\bibinfo {author} {\bibfnamefont {K.}~\bibnamefont
  {Pakdaman}}, \bibinfo {author} {\bibfnamefont {S.}~\bibnamefont {Tanabe}}, \
  and\ \bibinfo {author} {\bibfnamefont {T.}~\bibnamefont {Shimokawa}},\
  }\href@noop {} {\bibfield  {journal} {\bibinfo  {journal} {Neural Networks}\
  }\textbf {\bibinfo {volume} {14}},\ \bibinfo {pages} {895} (\bibinfo {year}
  {2001})}\BibitemShut {NoStop}%
\bibitem [{\citenamefont {Pikovsky}\ and\ \citenamefont
  {Kurths}(1997)}]{pikovsky1997coherence}%
  \BibitemOpen
  \bibfield  {author} {\bibinfo {author} {\bibfnamefont {A.~S.}\ \bibnamefont
  {Pikovsky}}\ and\ \bibinfo {author} {\bibfnamefont {J.}~\bibnamefont
  {Kurths}},\ }\href@noop {} {\bibfield  {journal} {\bibinfo  {journal} {Phys.
  Rev. Lett.}\ }\textbf {\bibinfo {volume} {78}},\ \bibinfo {pages} {775}
  (\bibinfo {year} {1997})}\BibitemShut {NoStop}%
\bibitem [{\citenamefont {Lindner}\ \emph {et~al.}(2002)\citenamefont
  {Lindner}, \citenamefont {Schimansky-Geier},\ and\ \citenamefont
  {Longtin}}]{lindner2002maximizing}%
  \BibitemOpen
  \bibfield  {author} {\bibinfo {author} {\bibfnamefont {B.}~\bibnamefont
  {Lindner}}, \bibinfo {author} {\bibfnamefont {L.}~\bibnamefont
  {Schimansky-Geier}}, \ and\ \bibinfo {author} {\bibfnamefont
  {A.}~\bibnamefont {Longtin}},\ }\href@noop {} {\bibfield  {journal} {\bibinfo
   {journal} {Phys. Rev. E}\ }\textbf {\bibinfo {volume} {66}},\ \bibinfo
  {pages} {031916} (\bibinfo {year} {2002})}\BibitemShut {NoStop}%
\bibitem [{\citenamefont {Qian}\ \emph {et~al.}(2000)\citenamefont {Qian},
  \citenamefont {Wang},\ and\ \citenamefont {Zhang}}]{qian2000stochastic}%
  \BibitemOpen
  \bibfield  {author} {\bibinfo {author} {\bibfnamefont {M.}~\bibnamefont
  {Qian}}, \bibinfo {author} {\bibfnamefont {G.-X.}\ \bibnamefont {Wang}}, \
  and\ \bibinfo {author} {\bibfnamefont {X.-J.}\ \bibnamefont {Zhang}},\ }\href
  {\doibase 10.1103/PhysRevE.62.6469} {\bibfield  {journal} {\bibinfo
  {journal} {Phys. Rev. E}\ }\textbf {\bibinfo {volume} {62}},\ \bibinfo
  {pages} {6469} (\bibinfo {year} {2000})}\BibitemShut {NoStop}%
\bibitem [{\citenamefont {Giacomelli}\ \emph {et~al.}(2000)\citenamefont
  {Giacomelli}, \citenamefont {Giudici}, \citenamefont {Balle},\ and\
  \citenamefont {Tredicce}}]{Giacomelli2000}%
  \BibitemOpen
  \bibfield  {author} {\bibinfo {author} {\bibfnamefont {G.}~\bibnamefont
  {Giacomelli}}, \bibinfo {author} {\bibfnamefont {M.}~\bibnamefont {Giudici}},
  \bibinfo {author} {\bibfnamefont {S.}~\bibnamefont {Balle}}, \ and\ \bibinfo
  {author} {\bibfnamefont {J.~R.}\ \bibnamefont {Tredicce}},\ }\href@noop {}
  {\bibfield  {journal} {\bibinfo  {journal} {Phys. Rev. Lett.}\ }\textbf
  {\bibinfo {volume} {84}},\ \bibinfo {pages} {3298} (\bibinfo {year}
  {2000})}\BibitemShut {NoStop}%
\bibitem [{\citenamefont {Ushakov}\ \emph {et~al.}(2005)\citenamefont
  {Ushakov}, \citenamefont {W\"unsche}, \citenamefont {Henneberger},
  \citenamefont {Khovanov}, \citenamefont {Schimansky-Geier}, ,\ and\
  \citenamefont {Zaks}}]{UshWun05}%
  \BibitemOpen
  \bibfield  {author} {\bibinfo {author} {\bibfnamefont {O.~V.}\ \bibnamefont
  {Ushakov}}, \bibinfo {author} {\bibfnamefont {H.-J.}\ \bibnamefont
  {W\"unsche}}, \bibinfo {author} {\bibfnamefont {F.}~\bibnamefont
  {Henneberger}}, \bibinfo {author} {\bibfnamefont {I.~A.}\ \bibnamefont
  {Khovanov}}, \bibinfo {author} {\bibfnamefont {L.}~\bibnamefont
  {Schimansky-Geier}}, , \ and\ \bibinfo {author} {\bibfnamefont {M.~A.}\
  \bibnamefont {Zaks}},\ }\href@noop {} {\bibfield  {journal} {\bibinfo
  {journal} {Phys. Rev. Lett.}\ }\textbf {\bibinfo {volume} {95}},\ \bibinfo
  {pages} {123903} (\bibinfo {year} {2005})}\BibitemShut {NoStop}%
\bibitem [{\citenamefont {Postnov}\ \emph {et~al.}(1999)\citenamefont
  {Postnov}, \citenamefont {Han}, \citenamefont {Yim},\ and\ \citenamefont
  {Sosnovtseva}}]{PosHan99}%
  \BibitemOpen
  \bibfield  {author} {\bibinfo {author} {\bibfnamefont {D.~E.}\ \bibnamefont
  {Postnov}}, \bibinfo {author} {\bibfnamefont {S.~K.}\ \bibnamefont {Han}},
  \bibinfo {author} {\bibfnamefont {T.~G.}\ \bibnamefont {Yim}}, \ and\
  \bibinfo {author} {\bibfnamefont {O.~V.}\ \bibnamefont {Sosnovtseva}},\
  }\href@noop {} {\bibfield  {journal} {\bibinfo  {journal} {Phys. Rev. E.}\
  }\textbf {\bibinfo {volume} {59}},\ \bibinfo {pages} {R3791} (\bibinfo {year}
  {1999})}\BibitemShut {NoStop}%
\bibitem [{\citenamefont {Miyakawa}\ and\ \citenamefont
  {Isikawa}(2002)}]{miyakawa2002experimental}%
  \BibitemOpen
  \bibfield  {author} {\bibinfo {author} {\bibfnamefont {K.}~\bibnamefont
  {Miyakawa}}\ and\ \bibinfo {author} {\bibfnamefont {H.}~\bibnamefont
  {Isikawa}},\ }\href@noop {} {\bibfield  {journal} {\bibinfo  {journal} {Phys.
  Rev. E}\ }\textbf {\bibinfo {volume} {66}},\ \bibinfo {pages} {046204}
  (\bibinfo {year} {2002})}\BibitemShut {NoStop}%
\bibitem [{\citenamefont {Kiss}\ \emph {et~al.}(2003)\citenamefont {Kiss},
  \citenamefont {Hudson}, \citenamefont {Santos},\ and\ \citenamefont
  {Parmananda}}]{kiss2003experiments}%
  \BibitemOpen
  \bibfield  {author} {\bibinfo {author} {\bibfnamefont {I.~Z.}\ \bibnamefont
  {Kiss}}, \bibinfo {author} {\bibfnamefont {J.~L.}\ \bibnamefont {Hudson}},
  \bibinfo {author} {\bibfnamefont {G.~J.~Escalera}\ \bibnamefont {Santos}}, \ and\
  \bibinfo {author} {\bibfnamefont {P.}~\bibnamefont {Parmananda}},\
  }\href@noop {} {\bibfield  {journal} {\bibinfo  {journal} {Phys. Rev. E}\
  }\textbf {\bibinfo {volume} {67}},\ \bibinfo {pages} {035201} (\bibinfo
  {year} {2003})}\BibitemShut {NoStop}%
\bibitem [{\citenamefont {Santos}\ \emph {et~al.}(2004)\citenamefont {Santos},
  \citenamefont {Rivera},\ and\ \citenamefont
  {Parmananda}}]{santos2004experimental}%
  \BibitemOpen
  \bibfield  {author} {\bibinfo {author} {\bibfnamefont {G.~J.~Escalera}\
  \bibnamefont {Santos}}, \bibinfo {author} {\bibfnamefont {M.}~\bibnamefont
  {Rivera}}, \ and\ \bibinfo {author} {\bibfnamefont {P.}~\bibnamefont
  {Parmananda}},\ }\href@noop {} {\bibfield  {journal} {\bibinfo  {journal}
  {Phys. Rev. Lett.}\ }\textbf {\bibinfo {volume} {92}},\ \bibinfo {pages}
  {230601} (\bibinfo {year} {2004})}\BibitemShut {NoStop}%
\bibitem [{\citenamefont {Vilela}\ and\ \citenamefont
  {Lindner}(2009{\natexlab{a}})}]{VilLin09}%
  \BibitemOpen
  \bibfield  {author} {\bibinfo {author} {\bibfnamefont {R.~D.}\ \bibnamefont
  {Vilela}}\ and\ \bibinfo {author} {\bibfnamefont {B.}~\bibnamefont
  {Lindner}},\ }\href@noop {} {\bibfield  {journal} {\bibinfo  {journal} {J.
  Theor. Biol.}\ }\textbf {\bibinfo {volume} {257}},\ \bibinfo {pages} {90}
  (\bibinfo {year} {2009}{\natexlab{a}})}\BibitemShut {NoStop}%
\bibitem [{\citenamefont {Vilela}\ and\ \citenamefont
  {Lindner}(2009{\natexlab{b}})}]{VilLin09b}%
  \BibitemOpen
  \bibfield  {author} {\bibinfo {author} {\bibfnamefont {R.~D.}\ \bibnamefont
  {Vilela}}\ and\ \bibinfo {author} {\bibfnamefont {B.}~\bibnamefont
  {Lindner}},\ }\href@noop {} {\bibfield  {journal} {\bibinfo  {journal} {Phys.
  Rev. E.}\ }\textbf {\bibinfo {volume} {80}},\ \bibinfo {pages} {031909}
  (\bibinfo {year} {2009}{\natexlab{b}})}\BibitemShut {NoStop}%
\bibitem [{\citenamefont {Lacasta}\ \emph {et~al.}(2002)\citenamefont
  {Lacasta}, \citenamefont {Sagu\'es},\ and\ \citenamefont
  {Sancho}}]{PhysRevE.66.045105}%
  \BibitemOpen
  \bibfield  {author} {\bibinfo {author} {\bibfnamefont {A.~M.}\ \bibnamefont
  {Lacasta}}, \bibinfo {author} {\bibfnamefont {F.}~\bibnamefont {Sagu\'es}}, \
  and\ \bibinfo {author} {\bibfnamefont {J.~M.}\ \bibnamefont {Sancho}},\
  }\href {\doibase 10.1103/PhysRevE.66.045105} {\bibfield  {journal} {\bibinfo
  {journal} {Phys. Rev. E}\ }\textbf {\bibinfo {volume} {66}},\ \bibinfo
  {pages} {045105} (\bibinfo {year} {2002})}\BibitemShut {NoStop}%
\bibitem [{\citenamefont {Sergeyev}\ \emph {et~al.}(2010)\citenamefont
  {Sergeyev}, \citenamefont {O’Mahoney}, \citenamefont {Popov},\ and\
  \citenamefont {Friberg}}]{sergeyev2010coherence}%
  \BibitemOpen
  \bibfield  {author} {\bibinfo {author} {\bibfnamefont {S.}~\bibnamefont
  {Sergeyev}}, \bibinfo {author} {\bibfnamefont {K.}~\bibnamefont
  {O’Mahoney}}, \bibinfo {author} {\bibfnamefont {S.}~\bibnamefont {Popov}},
  \ and\ \bibinfo {author} {\bibfnamefont {A.~T.}\ \bibnamefont {Friberg}},\
  }\href@noop {} {\bibfield  {journal} {\bibinfo  {journal} {Optics letters}\
  }\textbf {\bibinfo {volume} {35}},\ \bibinfo {pages} {3736} (\bibinfo {year}
  {2010})}\BibitemShut {NoStop}%
\bibitem [{\citenamefont {Cox}\ and\ \citenamefont
  {Lewis}(1966)}]{cox1966statistical}%
  \BibitemOpen
  \bibfield  {author} {\bibinfo {author} {\bibfnamefont {D.}~\bibnamefont
  {Cox}}\ and\ \bibinfo {author} {\bibfnamefont {P.}~\bibnamefont {Lewis}},\
  }\href@noop {} {\emph {\bibinfo {title} {The Statistical Analysis of Series
  of Events}}}\ (\bibinfo  {publisher} {John Wiley and Sons, New York},\
  \bibinfo {year} {1966})\BibitemShut {NoStop}%
\end{thebibliography}

%

\end{document}